\documentclass[12pt,letterpaper]{article}

\usepackage[includeheadfoot,
            marginratio={1:1,2:3}, 
            width=412pt, 
            height=688pt,]{geometry}

\usepackage{amsmath}
\usepackage{amsfonts}
\usepackage{amssymb}
\usepackage{stmaryrd}
\usepackage{graphicx}
\usepackage{tikz}
\usetikzlibrary{decorations.pathreplacing,calc}
\usepackage[activate={true,nocompatibility}]{microtype}

\usepackage{empheq}
\usepackage{paralist}
\usepackage{cite}
\usepackage{hyperref}

\newcommand{\nc}{\newcommand}
\nc{\lb}{\llbracket}
\nc{\rb}{\rrbracket}
\nc{\gl}{\llbracket}
\nc{\gr}{\rrbracket}

\newcommand{\eq}[1]{\begin{equation}
                     \begin{split} #1 \end{split}
                     \end{equation}}

\newcommand{\ov}{\overline}
\newcommand{\ovmod}[2][3]{{}\mkern#1mu\overline{\mkern-#1mu#2}}

\allowdisplaybreaks[2]
\numberwithin{equation}{section}
\begin{document}

\vspace*{-1.5cm}
\begin{flushright}
  {\small
  MPP-2019-21\\
%  LMU-ASC ??/17\\
  }
\end{flushright}

\vspace{1.5cm}
\begin{center}
{\LARGE
Swampland Variations on a Theme by KKLT \\[0.2cm]
} 
\vspace{0.4cm}

\end{center}

\vspace{0.35cm}
\begin{center}
  Ralph Blumenhagen, Daniel Kl\"awer and Lorenz Schlechter
\end{center}

\vspace{0.1cm}
\begin{center} 
\emph{Max-Planck-Institut f\"ur Physik (Werner-Heisenberg-Institut), \\ 
   F\"ohringer Ring 6,  80805 M\"unchen, Germany } \\[0.1cm]

\vspace{0.2cm}

 \vspace{0.3cm} 
\end{center} 

\vspace{0.5cm}

%%%%%%%%%%%%%%%%%%%%%%%%%%%%%%%%%%%%%%%%%%%%%%%
%%%%%%%%%%%%%%%%%%%%%%%%%%%%%%%%%%%%%%%%%%%%%%%
%%%%%%%%%%%%%%%%%%%%%%%%%%%%%%%%%%%%%%%%%%%%%%%
%%%%%%%%%%%%%%%%%%%%%%%%%%%%%%%%%%%%%%%%%%%%%%%
%%%%%%%%%%%%%%%%%%%%%%%%%%%%%%%%%%%%%%%%%%%%%%%
%%%%%%%%%%%%%%%%%%%%%%%%%%%%%%%%%%%%%%%%%%%%%%%
%%%%%%%%%%%%%%%%%%%%%%%%%%%%%%%%%%%%%%%%%%%%%%%
%%%%%%%%%%%%%%%%%%%%%%%%%%%%%%%%%%%%%%%%%%%%%%%

\begin{abstract}
The KKLT scenario in a warped throat, if consistent, provides a concrete
counterexample to both the AdS scale separation and the dS swampland 
conjectures.
First, we define and analyze the relevant effective field theory for
the conifold modulus and the overall
K\"ahler modulus that both  have exponentially small masses.  
The scalar potential still admits  KKLT-like AdS and dS minima.   
Second, we critically analyze the reliability of the employed  Wilsonian
effective action by evaluating  the masses of 
light modes localized in the warped throat.
The resulting mass spectrum is discussed with respect  to the
swampland distance conjecture. We find the recently observed emergent nature of the
latter not only at large distance points but also at the conifold
point motivating a general extension of it. In this respect, KKLT
  and trans-Planckian field distance are on equal footing.
It is pointed out that the reliability  of the KKLT minimum will
depend on how this emergent behavior is interpreted.
\end{abstract}

\clearpage

\tableofcontents

%%%%%%%%%%%%%%%%%%%%%%%%%%%%%%%%%%%%%%%%%%%%%%%
%%%%%%%%%%%%%%%%%%%%%%%%%%%%%%%%%%%%%%%%%%%%%%%
%%%%%%%%%%%%%%%%%%%%%%%%%%%%%%%%%%%%%%%%%%%%%%%
%%%%%%%%%%%%%%%%%%%%%%%%%%%%%%%%%%%%%%%%%%%%%%%
%%%%%%%%%%%%%%%%%%%%%%%%%%%%%%%%%%%%%%%%%%%%%%%
%%%%%%%%%%%%%%%%%%%%%%%%%%%%%%%%%%%%%%%%%%%%%%%
%%%%%%%%%%%%%%%%%%%%%%%%%%%%%%%%%%%%%%%%%%%%%%%
%%%%%%%%%%%%%%%%%%%%%%%%%%%%%%%%%%%%%%%%%%%%%%%

%\newpage

\section{Introduction}
\label{sec:intro}

The swampland program puts forward the idea that from the low-energy
perspective not every effective field theory does admit a UV
completion to a consistent theory of quantum gravity \cite{Vafa:2005ui}. Even though the
concrete framework of string compactifications with its built
in  formal constraints
has provided many examples of consistent effective low energy
theories, it turned out notoriously difficult to obtain certain
cosmologically or phenomenologically desired 
features like de Sitter vacua, large field inflation or length-scale
separation of a  large four-dimensional (observable) space-time and a
small compact six-dimensional space.
These short-comings have led to proposals for so-called swampland
conjectures that make an attempt to conceptually extract the
underlying quantum gravity reason why certain aspects cannot be
realized in the string landscape \cite{ArkaniHamed:2006dz,Ooguri:2006in,Klaewer:2016kiy,Ooguri:2016pdq,Obied:2018sgi,Cecotti:2018ufg,Klaewer:2018yxi}.

The weak gravity \cite{ArkaniHamed:2006dz} and distance conjectures
\cite{Ooguri:2006in} are two of the most studied and best understood
swampland criteria. The weak gravity conjecture (WGC) was originally
motivated by a semi-classical black hole decay argument, hinting at a
more profound reason behind it than simply circumstantial evidence
from special (perturbative) points in the moduli space of string
theory. A significant amount of effort has been put in the task of
providing convincing arguments for or even an explicit derivation of the
weak gravity conjecture, without using an explicit embedding into
string theory
\cite{Harlow:2015lma,Cottrell:2016bty,Hod:2017uqc,Fisher:2017dbc,Cottrell:2017ayj,Cheung:2018cwt,Hamada:2018dde,Urbano:2018kax,Montero:2018fns}. Over
the time it was realized that the different swampland conjectures are
not unrelated, but rather form a tight web with many interrelations
\cite{Klaewer:2016kiy,Palti:2017elp,Hebecker:2018vxz,Ooguri:2018wrx,Klaewer:2018yxi}. 

In fact, a modern point of view is that many of them can be explained by the emergence proposal, which states that weak couplings in the IR arise from integrating out an infinite tower of massive states which unitarize the theory in the UV \cite{Heidenreich:2017sim,Grimm:2018ohb,Heidenreich:2018kpg}. Of course, so far one only has a
finite amount of data and incomplete conceptual understanding so that such conjectures could eventually also
turn out to be too naive or even wrong. Because of this it is crucial to carefully scrutinize the 
available string theory evidence for and against them.

For example, in the case of the swampland distance conjecture (SDC) it
has been realized that it has to be refined in the sense that it
generically applies only for large scalar field displacements
$\Delta\Phi>\mathcal{O}(1)$ in Planck units
\cite{Baume:2016psm,Klaewer:2016kiy}. The distance conjecture and its
refinement have been subsequently tested in many settings
\cite{Cecotti:2015wqa,Palti:2015xra,Baume:2016psm,Klaewer:2016kiy,Valenzuela:2016yny,Blumenhagen:2017cxt,Palti:2017elp,Lust:2017wrl,Hebecker:2017lxm,Cicoli:2018tcq,Grimm:2018ohb,Heidenreich:2018kpg,Blumenhagen:2018nts,Blumenhagen:2018hsh,
  Lee:2018urn,Lee:2018spm,Grimm:2018cpv,Lee:2019tst}.

In this paper we also consider  two other swampland conjectures. The first one is
the (refined) dS swampland conjecture \cite{Obied:2018sgi,Ooguri:2018wrx} that states that\footnote{See also \cite{Andriot:2018wzk,Dvali:2018fqu,Garg:2018reu,Dvali:2018jhn,Andriot:2018mav} for other possible refinements of the dS swampland conjecture.}
\eq{
                  |\nabla V|\ge {c\over  M_{\rm pl}} \cdot V\,\qquad {\rm
                    or}\qquad {\rm min}(\nabla^2
                  V)\le -{c'\over M_{\rm pl}^2} \cdot V  
}
where $c, c'$ are of order one.
This conjecture in particular forbids de Sitter vacua.

The second one has not yet received the same kind of attention, but has
similar support as the dS swampland conjecture from tree-level supergravity 
compactifications \cite{Gautason:2015tig, Gautason:2018gln}. As also recently reviewed
in \cite{Danielsson:2018ztv},  it is notoriously difficult to 
find AdS$_4$ space-times that are truly four-dimensional in the sense
that one has a separation of mass-scales.
Thus, one can formulate an  {\it AdS scale separation
swampland conjecture} saying that AdS minima of string theory satisfy
\eq{
  m^2\, L^2_{\rm AdS}\le c''\,
}
 where $c''$ is an order one coefficient and $m$ is the lightest
 non-vanishing (moduli) mass.
For supersymmetric AdS$_4$
 vacua, there would be a holographic dual conjecture for
 three-dimensional conformal field theories \cite{deAlwis:2014wia}.
 
There is support
 for these conjectures at string tree level, though employing also
more involved quantum aspects of string theory, like $\alpha'$,
string-loop or even non-perturbative corrections, there are claims that
both conjectures can be falsified. The most famous example is the KKLT
scenario \cite{Kachru:2003aw}, which  is also the main topic of this paper.

To explain our motivation, let us recall some basic aspects of the KKLT
scenario and our motivation to propose a modified version of it.
Recall that the KKLT scenario provides a recipe for obtaining
 metastable de Sitter vacua that invokes  a
clever combination of classical and quantum effects. One considers type IIB compactifications
on warped Calabi-Yau spaces with non-trivial three-form fluxes.
In a first step, these fluxes stabilize the complex structure and the
axio-dilaton moduli, while the K\"ahler moduli remain massless.
The scalar potential is of no-scale type that admits Minkowski minima
which for $W_0\ne 0$ can break supersymmetry.

In a second step the K\"ahler modulus is stabilized by balancing 
a non-perturba\-tive effect against an exponentially small value of
$W_0$. Here one {\it assumes} that the string flux landscape does indeed admit
such small values of $W_0$. In this way one obtains a supersymmetric
AdS minimum that allows a separation of scales. Indeed,
one finds for the mass of the
K\"ahler modulus $ m_\tau  L_{\rm AdS}\sim a\tau\sim -\log W_0$ that, by  choice of a
exponentially  small value of $W_0$, can be made large.

In order to eventually get dS vacua, one uplifts the AdS-minimum by
the addition of an anti-D3-brane localized in a strongly warped
throat. It has been under debate whether this uplift mechanism is controlled (see \cite{Akrami:2018ylq} for a recent review).
In particular it was attempted to construct a Maldacena-Nu\~nez type no-go theorem \cite{Moritz:2017xto} for the KKLT uplift, which was strongly debated (see also \cite{Sethi:2017phn} for another criticism). The analysis of the 10D Einstein equations in this case depends on the details of implementing gaugino condensation on a D7-brane stack. Different approaches have led the authors of \cite{Gautason:2018gln,Gautason:2019jwq} to reconfirm the 10D analysis of \cite{Moritz:2017xto}, while the opposite conclusion was reached in \cite{Hamada:2019ack,Carta:2019rhx}. An attempt of implementing the flattening effects invalidating the uplift into the 4D picture \cite{Moritz:2017xto,Moritz:2018ani} was rejected by \cite{Kallosh:2018wme,Kallosh:2018psh,Kallosh:2019axr}.

As we will elaborate on,  we intend  to make a different point that
conceptually is prior to the uplift.
Our approach is related  in spirit to
\cite{Hebecker:2018yxs,Buratti:2018xjt}, where axions arising in 
warped compactifications were used to test and challenge the (0-form) weak gravity and distance conjectures.
Let us stress that it is important for the uplift to work that  one
has a highly warped throat in the first place so that the uplift can
be sufficiently tuned.
This means that in the first step the complex structure moduli have been stabilized such that
the Calabi-Yau indeed develops such a highly warped region, i.e.
that one is close to a conifold singularity.
As a consequence for stabilizing the complex structure moduli one
cannot work in the usual supergravity description, that, as explicitly
shown in \cite{Blumenhagen:2016bfp}, is only valid in the dilute flux limit. In this
limit,
the mass of the complex structure modulus $Z$ controlling the size of
the  three-cycle that shrinks to zero at the conifold singularity,
comes out as
\eq{
                   m_Z^2\sim {M_s^2\over {\cal V} |Z|^2}
} 
where ${\cal V}$ denotes the volume of the CY in units of the string
length ${\alpha'}$. Thus, for having the mass of this modulus to be
smaller
than the string scale, one needs ${\cal V} |Z|^2\gg 1$, which is the
dilute flux regime. 

Therefore, for KKLT one has to invoke an effective
action that is valid in the strongly warped regime, i.e. ${\cal V} |Z|^2\ll 1$.
This has been the subject of study
\cite{DeWolfe:2002nn,Giddings:2005ff,Frey:2006wv,Douglas:2007tu,Shiu:2008ry} already right after the seminal
paper \cite{Giddings:2001yu} by Giddings, Kachru, Polchinski (GKP).
Recently, the action of \cite{Douglas:2007tu} was employed in
\cite{Bena:2018fqc} to scrutinize the uplift mechanism in the KKLT
construction.
The main result of the  latter paper is that the uplift term strongly influences the
stabilization of $Z$ so that for too small quantized values of the
three-form flux the $Z$ modulus destabilizes. It is then a question
of the tadpole conditions whether sufficiently large fluxes can be
turned on\footnote{In view of the tadpole conditions for F-theory
compactifications
on CY fourfolds with a large Euler number $\chi=O(10^5)$, we view these
tadpole constraints not to be  too strong.}.

Another  result of \cite{Bena:2018fqc}
is that in the warped regime the mass of the  conifold  modulus is
hierarchically smaller than the mass of the bulk complex structure moduli.
However, then the question arises whether, in the second step, the
K\"ahler modulus can still be kept smaller than the conifold modulus.
Note that both of them are now exponentially light. 
It is the purpose
of this paper to study the Wilsonian effective theory for the conifold modulus
$Z$ and the overall K\"ahler modulus $T$ in the strongly warped
regime, thus continuing along the lines of \cite{Bena:2018fqc}.

The effective theory for these two very light moduli suggests an
alternative version of the KKLT scenario that differs from the
traditional one in the following aspect. Instead of assuming an
exponentially  small
value of $W_0$ in the landscape, in the first step we stabilize all additional
complex structure moduli and the axio-dilaton at a high scale in a
supersymmetric way.  Therefore, the effective theory for $Z$ and $T$ 
has $W_0=0$ and will be defined in the strongly warped regime, i.e. 
the periods of the CY will be expanded around the conifold point and the
warp factor will be taken into account.
As we will see, the KKLT scenario works in this case, as well. Both
moduli  get self-consistently stabilized in their ``perturbative''
regions $|Z|\ll 1$ and  ${\rm Re}(T)\gg 1$. Without the addition of 
anti D3-branes one finds AdS$_4$ minima with scale separation and with 
uplift term one gets meta-stable dS minima.  An exponentially small effective $W_0$ (for
K\"ahler modulus stabilization) will be dynamically generated by the stabilization of the conifold modulus.

Thus, this scenario seems to provide a concrete model of string moduli
stabilization in the highly non-classical regime, i.e.
\begin{itemize}
\item{close to a conifold singularity in the complex structure moduli
    space}
\item{including non-perturbative effects from stringy  D3-brane
    instantons or gaugino condensates on D7-branes}
\end{itemize}
that provides a counter-example against the two swampland conjectures
\begin{itemize}
\item{the (refined)   dS swampland conjecture forbidding dS minima}
\item{the AdS scale separation conjecture\,.}
\end{itemize}
To rescue those conjectures, one needs to find a loop-hole in the
computation or reject one of the assumptions being made. As mentioned, one mostly figured  that something is inconsistent with
the uplift mechanism.
However, this would only rescue the dS swampland conjecture. 

In the sections \ref{sec_two} and \ref{sec_three} we describe the
above mentioned modified KKLT construction.
In section \ref{sec_four} we quantitatively analyze a second potential inconsistency, namely that the used 
effective action for the strongly warped throat  might not be well controlled,
as there exist ultra-light Kaluza-Klein modes  that have a mass lighter
than the conifold modulus. 
We solve the Laplace
equation in the  warped throat by using both a simple  analytical approximation
and  a numerical  approach, where we are particularly  careful with the
dependence on the  relevant parameters of the model.
Both methods give consistent results indicating 
that there indeed exist eigenmodes that are supported in the
vicinity of the tip of the throat whose masses get highly redshifted so that
they become parametrically of the same  order as   the mass of the $Z$ modulus
itself. 

First reviewing and then applying the emergence hypothesis,
we will argue that the cut-off of the effective theory is not the
Planck scale but the mass scale of a D3-brane wrapping the three-cycle
that shrinks to a point at the conifold locus. 
Thus,  applying  the logic of the swampland distance conjecture to
the conifold locus we  find that, similar to infinite distance
points,  the metric on moduli space close to the conifold is
emerging by integrating out, in this case, a finite number of KK
modes.  This leads us to formulate an extension of the emergence
hypothesis of the swampland distance conjecture.
Two possible interpretations of this peculiar structure are discussed 
that lead to fairly opposite conclusions about the reliability  of the KKLT
scenario.
Either the utilized effective field theory is uncontrolled or
(one-loop)  quantum gravity effects from integrating out the tower of KK
modes are essentially harmless.

\section{A modified warped KKLT scenario}
\label{sec_two}

In this section, we introduce a slightly modified version of the KKLT
scenario that does not assume a landscape tuning of a tiny
$W_0$ in a non-supersymmetric minimum (though does not forbid 
such an extra tuning). Before we discuss the warped case, let us
review  moduli stabilization in the dilute flux limit.

\subsection{The conifold in the dilute flux limit}
\label{sec_two-one}

Let us consider type IIB (orientifold) compactifications on a Calabi-Yau threefold
${\cal M}$.
As usual, for stabilizing the complex structure and the axio-dilaton moduli,
we  turn on type IIB three-form fluxes. Since eventually we need to
design a warped throat we consider the region close to a conifold
singularity \cite{Candelas:1989js} in the complex structure moduli space.

There the threefold developes  a  nodal singularity, that
topologically can be considered as a cone over $S^2\times S^3$. The $S^3$ can
be made finite by deforming away from the conifold locus in the complex
structure moduli space, leading to the deformed conifold.
Call  $A$ the three-cycle that vanishes at the conifold and $B$ its
symplectic dual three-cycle. Then  the corresponding
periods $X^1=\int_A \Omega$, $F_1=\int_B \Omega$ have an expansion 
\eq{
\label{periods}
    \Pi=X^0\,\left( \begin{matrix}  1\\ Z\\ 
    -  {1\over 2\pi i} Z\log Z +C + DZ +  O(Z^2) \\
     \vdots \end{matrix}   \right)
}
with $\Pi^T=(X^0,X^1,F_1,\ldots)$ and  all other periods admitting a usual series expansion in $Z=X^1/X^0$.
The resulting  K\"ahler potential for the complex structure modulus $Z$
is given by
\eq{
\label{kahlerconi}
           K_{\rm cs}&=-\log\left[ -i \Pi^\dagger \Sigma\, \Pi \right]\\[0.1cm]
    &=-\log\left[ {1\over 2\pi} |Z|^2\, \log( |Z|^2) + A + O(|Z|^2)\right]
}
where $\Sigma$ is the symplectic pairing and $A>0$ is a real constant. 
The leading order K\"ahler metric reads 
\begin{equation}
  \label{eq:metricconi}
  G_{Z\ovmod Z}\sim -\log( |Z|^2)/A\;.
\end{equation}

\subsubsection*{Moduli stabilization}

\begin{figure}[ht]
  \centering
  \includegraphics[width=0.6\textwidth]{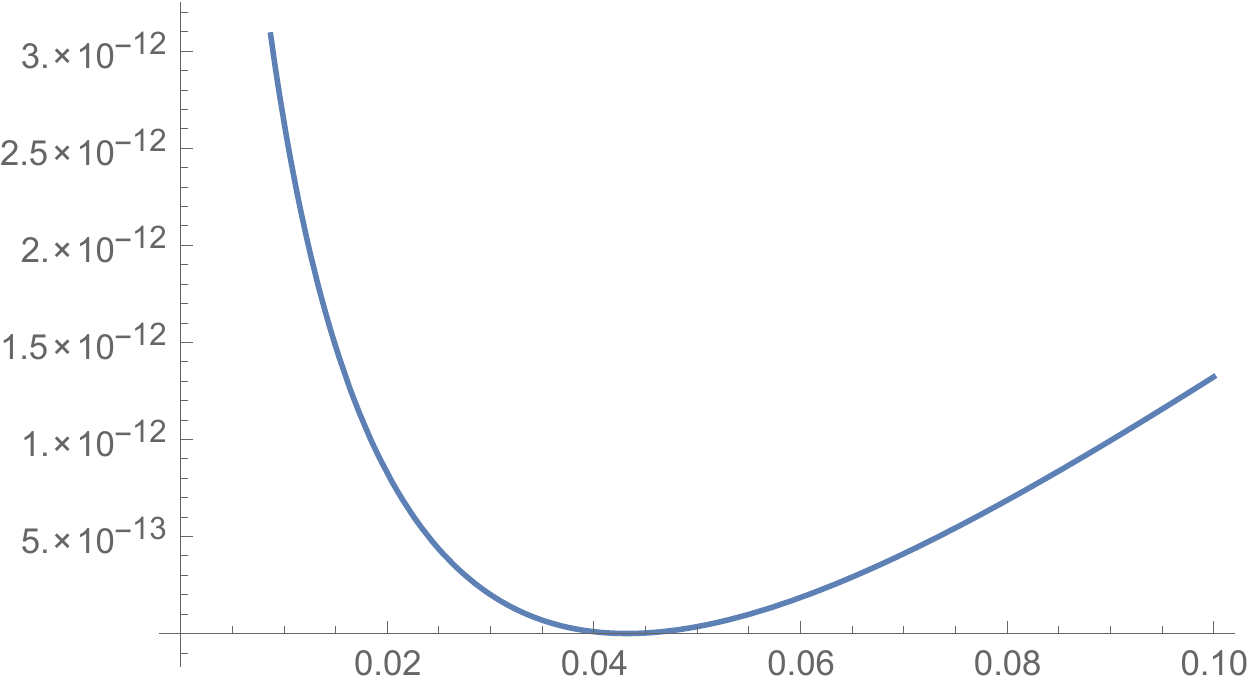}
\begin{picture}(0,0)
    \put(0,10){$\zeta$}
    \put(-219,141){$V$}
    \end{picture}
  \caption{Potential $V(\zeta)$ for $\sigma_0=0$ and the choice of
    parameters: $M=4$, $K=1$, $g_s=1/2$ and $\tau=10^3$.}
  \label{fig:0}
\end{figure}
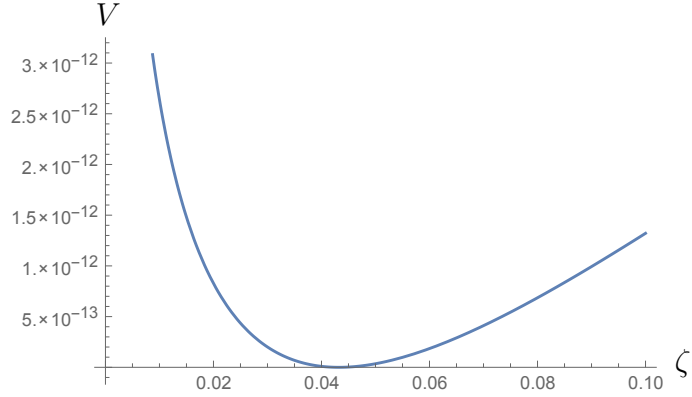

To stabilize the complex
structure moduli and the axio-dilaton $S=e^{-\varphi}+i C_0$ one turns
on NS-NS and R-R three-form fluxes.
The superpotential
generating the corresponding F-term scalar potential is of the familiar
Gukov-Vafa-Witten (GVW) type \cite{Taylor:1999ii,Gukov:1999ya}
\eq{
  \label{s_pot_02}
  W= \int_{\mathcal M} \bigl( F_3 +i S\, H_3 \bigr) \wedge \Omega_3\,
}
where $F_3=dC_2$ and $H_3=dB_2$ denote the R-R and NS-NS
three-form field strengths. Here the fluxes and the holomorphic three-form
are considered to be dimensionless, i.e. when cohomologically evaluated the
superpotential only depends on the periods and the flux quanta.
Taking also the K\"ahler potential 
$K=-3\log (T+\ovmod[1] T)-\log (S+\ovmod S)$
for the overall K\"ahler modulus
$T=\tau+i\theta$ and the axio-dilaton into account the resulting scalar potential 
\eq{
     V=e^K \left( G^{A\ovmod B} D_A W D_{\ovmod B} \ovmod[1] W -3 \vert W\vert^2
     \right)=e^K G^{I\ovmod J} D_I W D_{\ovmod J} \ovmod[1] W 
}  
is of no-scale type (here we have set $M_{\rm pl}=1$). Here $A,B$ run over all moduli and $I,J$ only
over the complex structure moduli and the axio-dilaton.
Now, let us turn on quantized $F_3$ form flux $M$ on the $A$-cycle and $H_3$ form
flux $K$ on the dual $B$-cycle so that 
the leading order dependence of the superpotential on the conifold
modulus $Z$ is like
\eq{
\label{superpot}
     W=-{M\over 2\pi i} Z (\log Z -1) +i  {K S} Z + \ldots \,.
}
Then the leading order contribution to the scalar potential becomes
\eq{
          V&\approx e^K \, G^{Z\ovmod Z} D_Z W D_{\ovmod Z} \ovmod[1] W\\
    &\approx  M^4_{\rm pl}\, {g_s \over \tau^3} \big[ -\log(|Z|^2)\big]^{-1} \,\Big\vert {M\over 2\pi}\log Z + K S\Big\vert^2\,
}
where in the second line we reintroduced the Planck-scale.
This potential
stabilizes the conifold modulus $Z=\zeta \exp(i\sigma)$ at 
\eq{
         \zeta_0=e^{-{2\pi K\over g_s M}}\,, \qquad \sigma_0=0
}
where for simplicity we assumed that the axio-dilaton is fixed at
$S=g_s^{-1}$. In figure  \ref{fig:0} we display the form of the
potential for $\sigma_0=0$.

\vspace{0.3cm}
%%%%%%%%%%%%
%%%%%%%%%%%%
%%%%%%%%%%%%

The mass (square) in the minimum are as usual given by the eigenvalues
of the matrix
$M^i{}_k=G^{ij} \partial_j\partial_k V$, which in this case can be shown to
scale as
\eq{
                   m^2_Z\sim {M_{\rm pl}^2\over {\cal V}^2 |Z|^2}\sim
                   {M_{\rm s}^2\over {\cal V} |Z|^2}\,
}
with ${\cal V}=\tau^{3\over 2}$.
Therefore, for the moduli masses to be smaller than the string scale,
one needs ${\cal V} |Z|^2\gg 1$, which is the so-called dilute flux
limit. In other words, the employed  supergravity action is only valid in
this limit, where the backreaction of the fluxes on the geometry can
be neglected. Note that the physical size of the three-cycle $A$ in
units of $\alpha'$ is given by
\eq{
             {\rm Vol}(A)={\cal V}^{1\over 2}\left|{\textstyle \int_A \Omega }\right|= ({\cal
               V} |Z|^2)^{1\over 2}\,
}
that makes the relevance of the combination $ {\cal V} |Z|^2$
manifest. As we will see, this combination will also appear very often
in the warped regime.

Let us note that the other complex structure moduli and the
axio-dilaton get masses that are lighter and scale as
\eq{
\label{masscso}
                    m^2_{\rm cs}\sim {M_{\rm pl}^2\over {\cal V}^2}\sim {M_{\rm s}^2\over {\cal V}}\,.
}

\subsubsection*{Wrapped D3-brane and emergence of the moduli space metric}

As was first observed in \cite{Strominger:1995cz}, the leading
divergence of the K\"ahler metric \eqref{eq:metricconi} can be
interpreted as an IR divergence from having integrated out the wrapped
D3-brane on the shrinking $S^3$ at the conifold locus which becomes
massless as we send $|Z|\to 0$. In general, integrating out a state
$\phi$ of
mass $m(Z)$ which depends on the modulus under consideration will
generate a one-loop contribution to its kinetic term, arising from the
trivalent vertex generated by expanding the mass term (of a boson)
\begin{equation}
  m^2(Z) \phi^2=m^2(Z_0) \phi^2+2m(Z_0)\frac{\partial m(Z)}{\partial Z}Z\phi^2+\dots\,.
\end{equation}
The result from integrating out $n_F$ fermions and $n_B$ bosons, including numerical factors is given by \cite{Grimm:2018ohb}
\begin{equation}
\label{eq:IntegrateOutBoseFermi}
  \delta g_{Z\ovmod Z}=\sum_{n=1}^{n_B}\frac{|\partial_Z
    m^B_n(Z)|^2}{8\pi^2}\left({\textstyle \frac{2\pi}{3\sqrt{3}}-1}\right)+\sum_{n=1}^{n_F}\frac{|\partial_Z m^F_n(Z)|^2}{8\pi^2}\log\left(\frac{\Lambda_{\rm UV}^2}{\left(m^F_n(Z)\right)^2}\right)\,.
\end{equation}
Due to supersymmetry, we are actually integrating out whole multiplets. The contribution from a chiral multiplet is
\begin{equation}
\label{eq:IntegrateOutSUSY}
  \delta g_{Z\ovmod Z}\sim\left|\partial_Z m(Z)\right|^2\left[1+\alpha\log\left(\frac{\Lambda_{\rm UV}^2}{\left(m(Z)\right)^2}\right)\right]\,,
\end{equation}
where $\alpha$ is an order one number. In the present case the state
that is integrated out in the effective field theory is the
hypermultiplet corresponding to the 
non-perturbative wrapped D3-brane on the shrinking $S^3$ of the
conifold.  Its mass is given by\footnote{Note that in our conventions
the volume ${\cal V}$ also contains a factor $g_s^{-3/2}$.}
\begin{equation}
    S_{D3}= \frac{1}{g_s}M_s^4\int_{\mathbb{R}\times
      S^3}\sqrt{-g}=\underbrace{\frac{1}{g_s^{1/4}}M_s({\cal
               V} |Z|^2)^{1\over 2}}_{m_{\rm D3}}\int_\mathbb{R} d\tau\,.
\end{equation}
Integrating out the corresponding multiplet with $\Lambda_{\rm
  UV}=M_{\rm pl}=M_s\mathcal{V}^{\tfrac12}/g_s^{1/4}$, for $|Z|\ll 1$ one obtains a dominant contribution from the fermions due to the large logarithm
\begin{equation}
\label{eq:IntegrateOutD3Unwarped}
   \delta g_{Z\ovmod Z}\sim -\log(|Z|^2)\,.
\end{equation}
This is precisely the  behavior of the tree level metric. The
emergence proposal says that in the field
theory  including the wrapped D3-brane hypermultiplet, 
the actual tree-level metric is non-singular and that the $\log$-term 
solely arises from integrating  out the light hypermultiplet.  

We will now include the flux induced warping in the discussion, focussing on the moduli stabilization. In section \ref{sec_four} we will come back to discuss the emergence of the K\"ahler metric also in the warped case.

\subsection{The conifold in the warped regime}
\label{sec_22}

As we have seen the  supergravity action from the last section is only
valid in the dilute flux regime where the backreaction of the fluxes
can be neglected. 
For really obtaining a strongly warped Calabi-Yau their  backreaction has
to be strong.
Indeed, it is well known that the backreaction of such a three-form flux and
of localized D3-branes on
the geometry leads to a warped CY metric \cite{Giddings:2001yu}
\eq{
\label{warpedmetr}
                 ds^2=e^{2A(y)} g_{\mu\nu} dx^\mu dx^\nu +
                            e^{-2A(y)} \tilde g_{mn} dy^m dy^n
}
where the warp factor $A(y)$ only depends on the internal coordinates
$y$ and $\tilde g_{mn}$ denotes the Ricci-flat metric on a CY
threefold. 
Locally an $H_3$ form flux on an $A$-cycle and an
$F_3$ form flux on its symplectic dual $B$-cycle leads to the warped
metric  on the deformed  conifold. This can be described as a cone over
$T^{1,1}$ cut off  in the IR by a finite size $S^3$.  This is the
Klebanov-Strassler (KS) solution \cite{Klebanov:2000hb}, whose metric is explicitly known
\eq{
\label{KSmetric}
    \widetilde{ds}^2={1\over 2} |S|^{2\over 3} K(y)\bigg[
     {dy^2+(g^5)^2\over 3 K^3(y)} &+ \cosh^2\left( {y\over2}\right)
    \left( (g^3)^2 + (g^4)^2\right) \\
    &+\sinh^2\left( {y\over2}\right)
    \left( (g^1)^2 + (g^2)^2\right) \bigg]
}
where $S$ is related to the conifold modulus $Z$ and
 the $g^i$ are a collection of  one-forms for the base
$S^2\times S^3$ and
\eq{
             K(y)={ (\sinh(2y)-2y)^{1\over 3}\over 2^{1\over 3}\, \sinh(y)}\,.
}
For $y=0$ the $S^2$ shrinks to zero size, while the $S^3$ stays
finite. Its  volume form can be read off from \eqref{KSmetric} as
$\omega_3\sim g^5\wedge g^3 \wedge g^4$.
The  warp factor of the KS solution reads
\eq{
\label{warpKS}
       e^{-4A(y)}=2^{2\over 3} {(\alpha' g_s M)^2\over |S|^{4\over 3}} {\cal
           I}(y)
}
where ${y}$ denotes the direction along the throat and 
\eq{
          {\cal
           I}({y})=\int_{y}^\infty dx {x \coth x -1\over
           \sinh^2 x} (\sinh (2x)-2x)^{1\over 3}\,.
}

We note that  the relation between the coordinates $S$ and $Z$ is a bit
more involved. This can be anticipated by noting that the coordinate
$Z$ is dimensionless while  the $S$ coordinate in the KS solution
  is not.  This is evident by the appearance of  $\alpha'$ in the warp
  factor \eqref{warpKS} implying that $S$ has dimension
  $[{\rm length}^3]$.
Recall that in the effective supergravity description
 there is no explicit dependence on the string scale, those factors
 are absorbed in the definition of the superfields and $M_{\rm pl}$.

Moreover,  scaling the internal metric via $\tilde g\to \lambda^2 \tilde g$
describes the breathing mode of the CY, i.e. the K\"ahler modulus 
for the overall volume. As the fluxes do not stabilize the
K\"ahler moduli, this should better be an unconstrained deformation.
There exists the relation  $\lambda\sim {\cal V}_w^{1/6}$ 
where\footnote{Our notation is related to the one used in
  \cite{Douglas:2007tu, Bena:2018fqc} by a  rescaling of the moduli fields $\tau$ and $Z$
  (called $\rho$ and $S$ in \cite{Douglas:2007tu, Bena:2018fqc}) by suitable powers  of
  $||\Omega||^2$ and $V_w$. Note that the latter two quantities 
 are not considered to be moduli dependent  but just values
 around which one expands.}
\eq{
{\cal V}_w= {1\over g_s^{3/2} (\alpha')^{3}} \int d^6y\, e^{-4A} \sqrt{
    \tilde g}\sim \tau^{3\over 2}
}
denotes the warped volume of the CY in units of $\alpha'$.
In \cite{Giddings:2005ff}  it was shown that the 10D string equations of motion admit
an unconstrained deformation $\lambda$ only if the warp factor
scales non-trivially
\eq{
\label{warpfactor1}
           e^{-4A}=1+ {e^{-4A_{\rm  con}}\over \lambda^4}\sim 1+
           {c\over  ({\cal V}_w |Z|^2)^{2\over 3}}+\cdot
\,.
}
where we have chosen the warp factor to be one in the large volume,
unwarped regime. Putting the last two observations together, 
the coordinate $S$ in the KS solution
 \eqref{KSmetric}, \eqref{warpKS} and the conifold coordinate $Z$ are related via the 
rescaling  
\eq{
\label{rescale}
S\to (\alpha')^{3/ 2} \sqrt{g_s^{3/2} \,{\cal V}_w} Z\,.
}
Then  one can
 write the warp factor close to the conifold locus as
\eq{
\label{warpfactor2}
              e^{-4A({y})}\approx 2^{2\over 3} {g_s M^2\over
                ({\cal V}_w |Z|^2)^{2\over 3}} {\cal
           I}(y) \,. 
}
Even though we were not very careful with numerical prefactors, this
is the relation we will
use in the following. 
Therefore, the regime of strong warping is given by ${\cal V}_w
|Z|^2\ll 1$.  We notice that it makes manifest the orthogonality of the complex
structure and K\"ahler moduli even in the warped case. Keeping $Z$
fixed and close to the conifold locus, by scaling up the metric one
can reach the dilute flux regime where the warp factor goes to one.

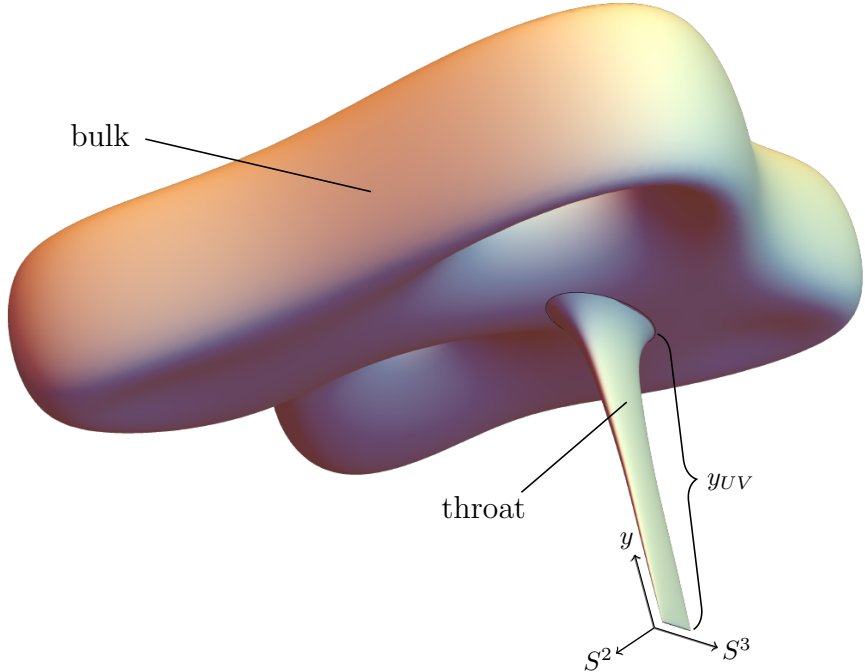
\begin{figure}[ht]
%\centering
\vspace{-2ex}
  		\begin{tikzpicture}
  		\tikzset{
    			shadowY/.style={preaction={transform canvas={shift={(-0.3pt,-0.1pt)}},draw=gray,semithick}},
    			shadowS2/.style={preaction={transform canvas={shift={(-0.1pt,-.1pt)}},draw=gray,semithick}},
    			shadowS3/.style={preaction={transform canvas={shift={(0pt,-0.3pt)}},draw=gray,semithick}},
 				}
    	\node[anchor=south west,inner sep=0] (modulispace) at (0,0) {\includegraphics[width=0.8\textwidth]{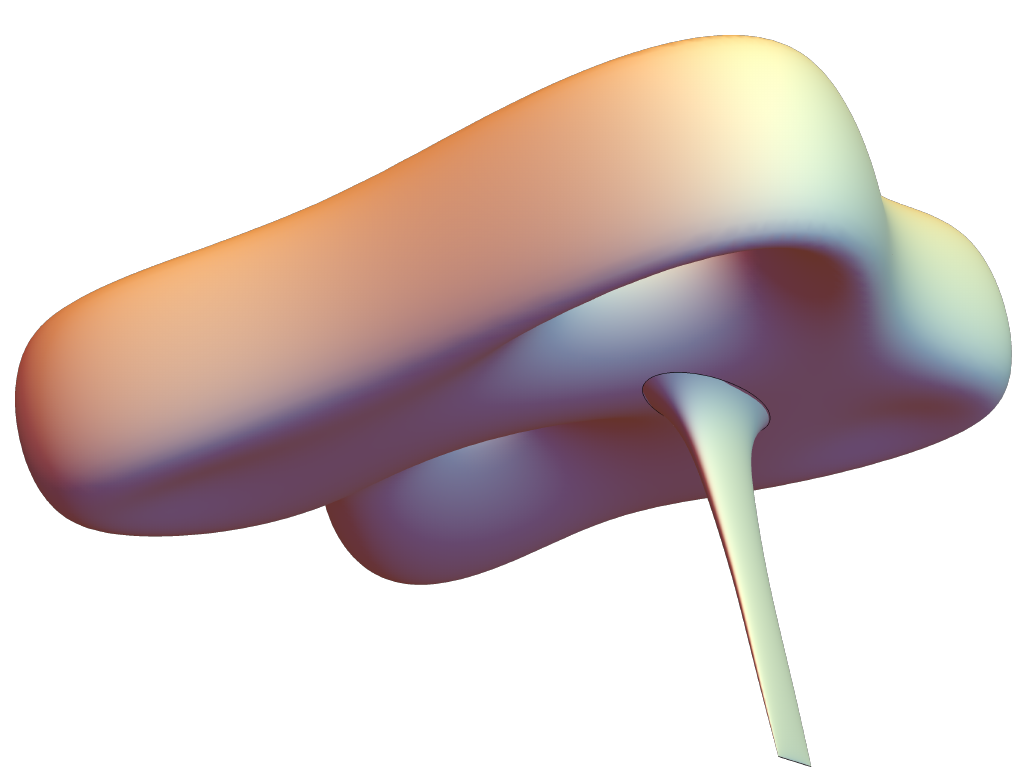}};
    	%coordinate system
		\draw[->,shadowY] (8.77,0.2) -- ($ (8.75,0.2) + (104:1) $);%y
		\draw[->,shadowS2] (8.77,0.2) -- ($ (8.75,0.2) + (215:0.6) $);%S2
		\draw[->,shadowS3] (8.77,0.2) -- ($ (8.75,0.2) + (-17:0.9) $);%S3
		%bulk/throat arrows
		\draw[-,semithick] (2,6.7) -- (5,6);%Bulk
		\draw[-,semithick] (7,2) -- (8.4,3.2);%Throat
		%coordinatelabels
		\node(y) at ($ (8.75,0.2) + (107:1.2) $) {\footnotesize $y$};
		\node(S2) at ($ (8.75,0.2) + (210:0.84) $) {\footnotesize $S^2$};
		\node(S3) at ($ (8.75,0.2) + (-14:1.15) $) {\footnotesize $S^3$}; 
		%bulk/throat labels 
		\node(bulk) at (1.4,6.75) {bulk};  
		\node(throat) at (6.5,1.8) {throat};
		%curly brace
		\draw [decorate,decoration={brace,amplitude=8pt},semithick,xshift=-4pt,yshift=0pt]
		($ (8.92,4.1) + (10:0.03) $) -- ($ (9.4,0.17) + (10:0.03) $) node [black,midway,xshift=21pt] {\footnotesize
		$y_{UV}$};
		\end{tikzpicture}
	\caption{A sketch of a Calabi-Yau with a KS-throat. At the tip of the throat the $S^2$ shrinks to zero size while the $S^3$ remains finite. $y_{UV}$	marks the cutoff where the throat meets the bulk.} 
	\label{fig_throat}
\end{figure}

In the strongly warped regime close to a conifold singularity in
complex structure moduli space, the CY develops a long throat region
that locally can be described by a KS throat and which is glued at a
UV scale $y_{\rm UV}$ to the remaining bulk threefold. This picture
of the CY is pictorially shown in figure \ref{fig_throat}.

\subsubsection*{Bounds on parameters}

Let us derive  lower bounds on the parameters $(g_s, M)$ and $y_{\rm
  UV}$ resulting from the suppression of $\alpha'$ corrections.
First, in order for the supergravity, large radius description to be consistent
one requires that the  size of the $S^3$ at the tip of the
conifold stays larger than the string length.  This can be simply
read-off  from the warped KS metric  \eqref{warpedmetr} and \eqref{KSmetric}
\eq{
       R^2_{S^3}\sim e^{-2A(0)} |S|^{2\over 3} \sim \alpha' g_s |M|
}
where we used the substitution \eqref{rescale}. Therefore, throughout
the following we work in the regime
\eq{
\label{sugraregime}
                         1\ll g_s |M| \,.
}
Similarly, one can derive a lower bound on $y_{\rm UV}$ by demanding
that the proper field length of the KS throat 
\eq{
              L_{\rm throat}= \int_0^{y_{\rm UV}}  dy \sqrt{G_{yy}} 
}
measured with the warped metric
is larger than the string length. Using the warped KS metric we find
\eq{
        L_{\rm throat}&\sim \big(\alpha' g_s |M|\big)^{1\over 2}  \int_0^{y_{\rm UV}}  dy\,
        {{\cal I}^{1\over 4}(y)\over K(y)}\sim \big(\alpha' g_s |M|\big)^{1\over 2}\, y_{\rm UV}
}
where we expanded the integrand for small values of $y$ as 
${{\cal I}^{1/4}(y)/K(y)}\approx 1.33 + O(y^2)$. Therefore, we obtain the
lower bound\footnote{Later we will often encounter the combinations 
$g_s M^2 \gg |M|$ and $g_s M^2\,  y_{\rm UV}^2 \gg |M|$.}
\eq{
\label{lowboundy}   
                  1\ll g_s |M|\,  y_{\rm UV}^2 \,.
}
As already practiced, we will not be concerned with order one numerical
prefactors, but will be very careful with the dependence of various
quantities on the parameters
\eq{
                \big\{ g_s, M, {\cal V}_w, Z, y_{\rm UV} \big\}\,. 
}

In \cite{Douglas:2007tu} (for a  recent summary see also
\cite{Bena:2018fqc}) an effective action was proposed that describes
the dynamics of the $Z$ modulus and the overall K\"ahler modulus in
the region of the $Z$ moduli space close to the conifold singularity.
This is the action that we will further study in this paper, as it is
the one underlying the warped KKLT scenario. Let us start with the
K\"ahler potential.

\subsubsection*{The modified no-scale structure}

Doing the same substitution  \eqref{rescale} 
 for the K\"ahler metric and the corresponding K\"ahler potential
in \cite{Douglas:2007tu, Bena:2018fqc}, we obtain in the
strongly warped regime\footnote{By order $O(\xi)$
  we actually mean first order in ${\xi \vert Z\vert^{2\over
      3}\over (T+\ovmod[1] T)}$.}
\eq{ 
\label{kaehlerpotb}
    K=-3\log(T+\ovmod[1] T) + {c' \xi \vert Z\vert^{2\over
      3}\over (T+\ovmod[1] T)} + O(\xi^2)
} 
with $\xi= g_s M^2$ and $c'$ an order one numerical factor whose
precise value  we take \cite{Bena:2018fqc} as   $c'\approx 1.18$.

Since the second term mixes the $T$ and the $Z$ moduli, there will be
off-diagonal terms in the K\"ahler metric. However, just turning on
three-form
flux should not stabilize the K\"ahler moduli. Therefore, something
non-trivial must happen to preserve the no-scale structure of the
induced scalar potential. To make this evident we  
consider the following more general ansatz for the K\"ahler potential
\eq{ 
    K=-3\log(T+\ovmod[1] T) + {c'\xi \vert Z\vert^{2\over
      3}\over (T+\ovmod[1] T)^N} + O(\xi^2)\,
} 
where we leave the exponent $N$ open for the moment. 
Then it is straightforward to compute 
\eq{
\label{noscalea}
    \sum_{I,\ovmod J} G^{I\ovmod J} \partial_I K \partial_{\ovmod J} K = 3-(N-1)
    {c'\xi \vert Z\vert^{2\over
      3}\over (T+\ovmod[1] T)^N} + O(\xi^2) 
}
where the sum runs over the set $I,J\in\{T,Z\}$.
Therefore, we see that for $N=1$ the linear term in $\xi$ precisely
vanishes. This is nothing else than the expected no-scale structure 
for the warped CY case. 
Similarly, one can show that
\eq{
\label{noscaleb}
        G^{Z \ovmod J}\partial_{\ovmod J} K =3(1-N) Z + O(\xi)
}
so that also for this combination the leading order correction
cancels precisely for $N=1$. 

\subsubsection*{Stabilization of conifold modulus}

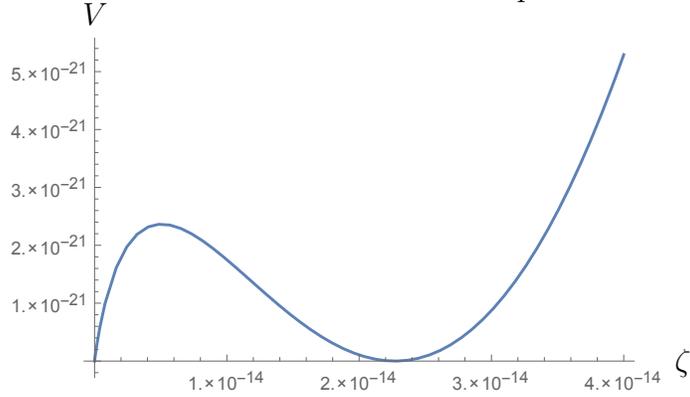
\begin{figure}[ht]
  \centering
  \includegraphics[width=0.6\textwidth]{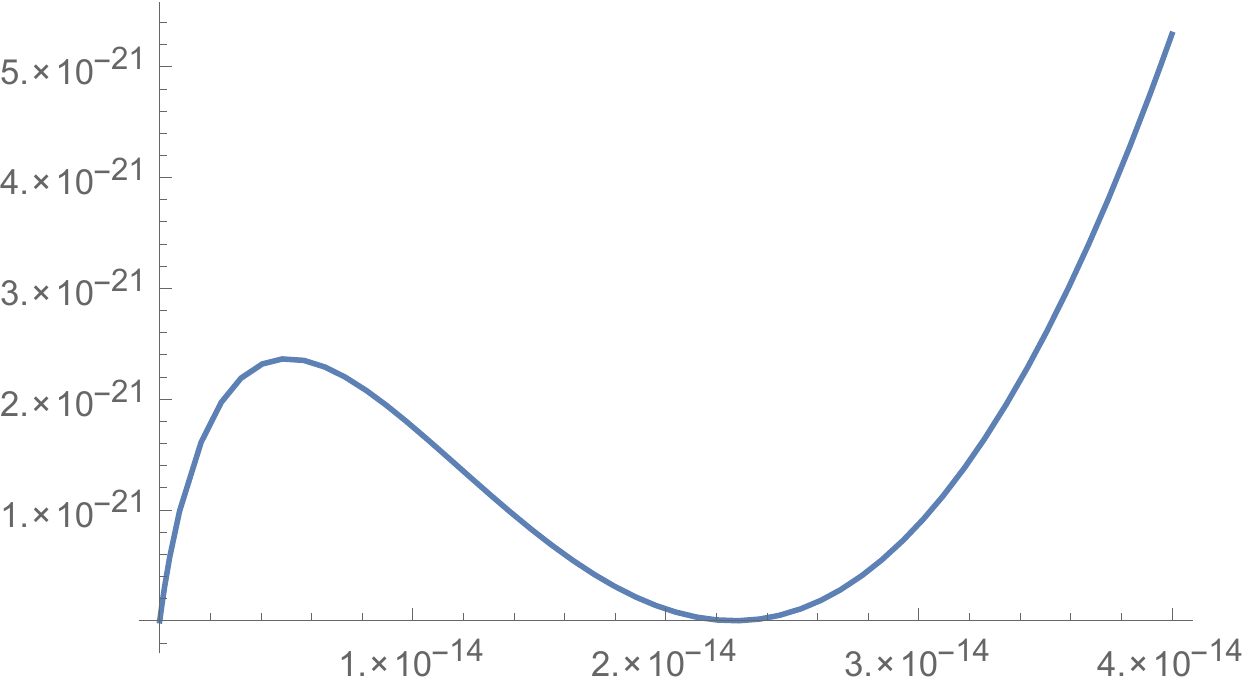}
\begin{picture}(0,0)
    \put(0,10){$\zeta$}
    \put(-224,141){$V$}
    \end{picture}
  \caption{Potential $V(\zeta)$ for $\sigma_0=0$ and the choice of
    parameters: $M=1$, $K=1$, $g_s=1/5$ and $\tau=10$.}
  \label{fig:1}
\end{figure}

Now, by turning on three-form flux supported on 3-cycles in the
bulk of the CY threefold, all remaining
complex structure moduli and the axio-dilaton
can be  stabilized and have a mass scale as in \eqref{masscso}.
The conifold modulus $Z$ is expected to be much lighter so  that the
former moduli  can  be integrated out. 
Therefore,  we are considering an effective theory for the
two moduli, $Z$ and $T$. Say after the first step the superpotential
in the minimum takes the value $W_{\rm cs}$ so that by again turning
on $F_3$ form flux on the $A$-cycle and $H_3$ form flux on the $B$-cycle
the total superpotential becomes
\eq{
\label{superpotb}
     W=W_{\rm cs}-{M\over 2\pi i} Z (\log Z -1) +i  {K S} Z  \,,
}
where assuming $Z\ll 1$ we neglected the higher order terms in the corresponding periods.
Due to the warped no-scale structure the resulting scalar potential
\eq{
     V=e^K \left( G^{I\ovmod J} D_I W D_{\ovmod J} \ovmod[1] W -3 \vert W\vert^2
     \right)
}  
simplifies considerably 
\eq{
\label{scalarpotno}
          V&= e^K \, G^{Z\ovmod Z} \partial_Z W \partial_{\ovmod Z} \ovmod[1] W\\
    &= {18 g_s\over c'\xi} {\vert Z\vert^{4\over 3}\over (T+\ovmod[1] T)^2} \Big\vert
      {M\over 2\pi}\log Z + K S\Big\vert^2\\
    &= {9\over 2 c' M^2} {\zeta^{4\over 3}\over \tau^2} \Big[
      \Big({\textstyle {M\over 2\pi}}\log \zeta+{\textstyle {K\over g_s}}\Big)^2 +
      \big({\textstyle {M\over 2\pi}}\big)^2 \,\sigma^2 \Big]\,.
}
This scalar potential is positive definite and vanishes for
$\partial_Z W=0$ which gives
\eq{
         \zeta_0=e^{-{2\pi K\over g_s M}}\,, \qquad \sigma_0=0\,.
}
Note that due to the no-scale relation \eqref{noscaleb} and 
\eq{
\label{noscalec}
               G^{T\ovmod Z} \partial_{\ovmod Z}K + G^{T\ovmod[1] T} \partial_{\ovmod[1]
                 T}K=-2\tau +{c' \xi \over 3} |Z|^{2\over
                 3}+O(\xi^2)\approx -2\tau\,,
} 
we obtain (up to the order to which we compute)
\eq{   
             F^Z&=G^{Z\ovmod Z}(\partial_{\ovmod Z} \ovmod[1] W +\partial_{\ovmod Z} K\,\ovmod[1] W)
             +G^{Z\ovmod[1] T}\partial_{\ovmod[1] T} K\,\ovmod[1] W \approx 0\,,\\
              F^T&=G^{T\ovmod Z} \partial_{\ovmod Z} K\,\ovmod[1] W
             +G^{T\ovmod[1] T}\partial_{\ovmod[1] T} K\,\ovmod[1] W \approx -2\tau\ovmod[1] W\,
}
so that  supersymmetry breaking can only occur along the K\"ahler moduli.
In figure \ref{fig:1} we display the scalar potential as a function
of the conifold modulus $\zeta$.  As already observed in \cite{Bena:2018fqc}, away
from the vicinity of the Minkowski minimum,
it shows a different functional behavior than the unwarped case.
%\vspace{0.3cm}

%%%%%%%%%%%%
%%%%%%%%%%%%
%%%%%%%%%%%%

%\noindent
The value of the superpotential in the Minkowski minimum is
given by
\eq{      
\label{w0effi}
     W_0=W_{\rm cs}-{M\over 2\pi i}  Z_0 = W_{\rm cs}-{M\over 2\pi i}\, e^{-{2\pi K\over
         g_s M}} \,.
}
Note that the second term is exponentially small. For getting a
supersymmetric minimum, i.e. a warped CY threefold, the value $W_{\rm
  cs}$ after fixing the heavy moduli, have to be such that it cancels
the second term. In general though supersymmetry is broken and by
having already $W_{\rm cs}=0$ in the first step, the scale of
breaking
is exponentially small. In the following we investigate this special
case, even though our analysis also goes through for more general
(small) values of $W_{\rm cs}=0$.

Using the scalar potential \eqref{scalarpotno} and the K\"ahler metric
following from the K\"ahler potential \eqref{kaehlerpotb}, the masses of both two
real components  of the complex structure modulus $Z$ scale as
\eq{
\label{massconi}
                m^2_Z\simeq  {\big({\cal V}_w \vert
                  Z\vert^2\big)^{1\over 3}\over  g_s M^2}
                  {M_{\rm pl}^2\over {\cal V}_w} \simeq   {\big({\cal
                      V}_w \vert Z\vert^2\big)^{1\over 3}\over g_s^{3/2}\, M^2}
                  M_{\rm s}^2 \,
}
so that in the strong warping regime  $ {\cal V}_w \vert Z\vert^2\ll 1$
they are indeed smaller than
the string scale.  Moreover, the mass of the $Z$ modulus is
exponentially
smaller than the mass \eqref{masscso} of the other complex structure moduli and the
axio-dilaton, justifying they could be  integrated out in a first step.

\subsection{Anti D3-brane uplift}

Adding a single  $\ov{\rm D3}$-brane at a position $y$ in the warped
throat induces
an extra contribution to the scalar potential \cite{Kachru:2002gs,Kachru:2003aw}
\eq{
            S_{\ov{\rm D3}}\sim 2\, {M_s^4\over g_s} \int d^4 x \sqrt{-g}\,  e^{4A({y})} \sim \int
            d^4x  \sqrt{-g}\ \, {2 \,M_{\rm pl}^4 \over \tau^3} e^{4A({y})} 
}
Using the warp factor \eqref{warpfactor2} and that the scalar
potential has its minimum at the tip of the throat ${y}=0$, the scalar potential of the $\ov{\rm
  D3}$-branes can be written as
\eq{
\label{Vantid3}
          V_{\ov{\rm D3}}= {9 c'' \over 2 g_s M^2} {\zeta^{4\over 3}\over \tau^2} 
}
with an order one coefficient $c''$, which  we take  from
\cite{Bena:2018fqc} to be  $c''=2^{1\over 3}/{\cal I}(0)\approx 1.75$.
Note that only with the factor ${\cal V}_w^{2/3}$ in the warp
factor \eqref{warpfactor1}, we get the correct and always used $V_{\ov{\rm
    D3}}\sim\tau^{-2}$ behavior of the uplift potential for  an
$\ov{\rm D3}$-brane at the tip of a strongly warped throat.
The total potential is then given by
\eq{
\label{Vantitot}
V_{\rm tot}= {9\over 2 c' M^2} {\zeta^{4\over 3}\over \tau^2} \Big[
      \Big({\textstyle {M\over 2\pi}}\log \zeta+{\textstyle {K\over g_s}}\Big)^2 +
      {c' c''\over g_s} \Big]\,
}
where we minimized already the $\sigma=\arg(Z)$ modulus.
%\footnote{Note  that the scalar potential \eqref{Vantitot} differs slightly by a
%  factor of $g_s^{-1}$ in the last term from the result given in \cite{Bena:2018fqc}.}.
As already noticed in \cite{Bena:2018fqc}, both the flux and the $\ov{\rm D3}$
brane contribution scale like ${\zeta^{4/3}\over \tau^2}$ so that
for the stabilization of $\zeta$ the uplift cannot be neglected, as it
is usually done for KKLT.
The extrema of the for $\zeta>0$ positive definite total potential $V_{\rm tot}$ were determined analytically in \cite{Bena:2018fqc}
\eq{
        \zeta=\exp\bigg({\textstyle -{2\pi\over g_s}{K\over M} -{3\over 4}\pm \sqrt{{9\over
            16}-{4\pi^2\over g_s M^2} c' c''} }\bigg)
}
where the positive sign corresponds to a local minimum and the
negative sign to a local maximum.

It was observed in \cite{Bena:2018fqc} that the extrema are gone if
$g_s M^2\le 12^2$ and
then it becomes a question of the tadpole cancellation condition whether
sufficiently large fluxes are allowed. This issue could potentially
spoil the uplift and therefore the validity of the KKLT scenario.
De Sitter vacua could not be achievable any more, while supersymmetric 
AdS$_4$ compactification with scale separation would still be
possible. In view of large tadpole constraints appearing in F-theory
compactifications on Calabi-Yau fourfolds, in the following, we do not consider this
issue to be really severe and assume that sufficiently large fluxes
can be turned on.

\section{Stabilization of K\"ahler modulus}
\label{sec_three}

So far the potential still features a modified no-scale structure so
that the overall K\"ahler modulus $T$ remains as a flat direction.
In KKLT this modulus is stabilized by non-perturbative effects coming
either from Euclidean D3-brane instantons or gaugino condensation on
D7-branes. The employed effective superpotential after integrating out the
complex structure moduli and the axio-dilaton reads
\eq{
              W=W_0 + A\, e^{-a T}
}
where the one-loop Pfaffian $A$ can only depend on the complex
structure moduli and can therefore be considered to be constant.
If one assumes that $W_0$ can be tuned exponentially small in the
string
landscape, then the two terms in $W$ can compete with each other and
one finds a supersymmetric AdS$_4$ minimum at $\tau \exp(-a \tau)\sim |W_0|$
so that the mass of $\tau$ scales as $m^2_\tau\sim -|W_0|^2/\log W_0$ which is
also exponentially small.

However, as we have seen the mass of the $Z$ modulus is also
exponentially small so that it is a priori not clear that one is
allowed to integrate $Z$ out before stabilizing $T$. Therefore, we move
one step back and consider a supergravity model with two moduli, $Z$
and $T$, and look for minima of the full scalar potential.
This model is specified by the K\"ahler potential
\eq{ 
    K=-3\log(T+\ovmod[1] T) + {c'\xi \vert Z\vert^{2\over
      3}\over (T+\ovmod[1] T)} + O(\xi^2)\,
} 
and the KKLT-like superpotential
\eq{
          W=W^{(Z)}+W^{(T)}=-{M\over 2\pi i} Z (\log Z -1) +i  {K\over g_s} Z + A e^{-aT}
}
where we have set $W_{\rm cs}=0$ and will  assume that the Pfaffian
$A>0$ is real, positive and does not depend on the conifold modulus $Z$.

Let us analyze analytically the form of the resulting total scalar
potential. Due to the no-scale relations \eqref{noscalea},
\eqref{noscaleb} and \eqref{noscalec}
it simplifies considerably
\eq{
        V\approx e^K\left( G^{I\ovmod J} \partial_I W \partial_{\ovmod J}
          \ovmod[1] W
          -2\tau\Big(\partial_T W^{(T)} \ovmod[1] W + W \partial_{\ovmod[1] T}\ov
          W^{(T)}\Big)\right)\,.
}
If the conifold modulus $Z$ is still stabilized by $\partial_Z
W^{(Z)}=0$ at $Z_0$ then the resulting potential for the K\"ahler
modulus will become
\eq{
          V\approx  e^K\left( G^{T\ovmod[1] T} \partial_T W^{(T)} \partial_{\ovmod[1] T}
          \ovmod[1] W^{(T)}
          -2\tau\Big(\partial_T W^{(T)} \ovmod[1] W + W \partial_{\ovmod[1] T}\ov
          W^{(T)}\Big)\right)
}
with $W=W_0+W^{(T)}$ and $W_0=W^{(Z)}(Z_0)$. This is nothing else
than the usual KKLT scalar potential with an exponentially small $W_0=-{M\over
  2\pi i} \exp(-{2\pi\over g_s}{K\over M})$. Therefore, the minimum
will be supersymmetric with the gravitino mass
\eq{
            m_{3/2}=e^{K/2} |W| \sim { g_s^{1/2} M |Z_0|\over (4\pi) \tau_0^{3/2}} M_{\rm pl}
}
and the value of the scalar potential in the AdS minimum
\eq{
\label{AdScosmo}
          V_0\sim -m_{3/2}^2\, M_{\rm pl}^2 \sim -{ g_s M^2 |Z_0|^2\over 16\pi^2\tau^3_0} M_{\rm pl}^4\,.
}

In order for this two-step procedure to be self-consistent, one needs
that eventually the K\"ahler modulus is much lighter than the conifold modulus. 
Let us estimate their masses.
As we have seen, the mass of $Z$ scales at
\eq{
                            m_Z^2\sim {|Z_0|^{2\over 3}\over g_s M^2 \,\tau_0}
                            M_{\rm pl}^2
}
while the KKLT scenario fixes the mass of the K\"ahler modulus at
\eq{
             m_\tau^2\sim {a^2 |W_0|^2\over \tau_0} M_{\rm pl}^2 \sim
                      {a^2 M^2\, |Z_0|^2\over \tau_0} M_{\rm pl}^2
}
so their ratio is
\eq{   
 {m^2_\tau\over m^2_Z}\sim  \left(M^3\, |Z_0|\right)^{4\over 3}\ll 1\,.
}
Here we have also taken into account the powers of the flux quantum
$M$ (as this can be large).
This analysis suggests that the minimum of the total scalar potential
is given at 
\eq{
\label{theomods}
        {\rm no-scale\ minimum:}\quad  \partial_Z W^{(Z)}=0\,\Longrightarrow  \zeta=e^{-{2\pi\over
             g_s}{K\over M}}\,,\    \   \sigma=0\\[0.1cm]
    {\rm KKLT\ minimum:} \quad A(2a\tau +3 )- 3|W_0| e^{a\tau}=0\,,\  \ \theta=-{\pi/2}  \,.                 
}
In the following we will consider concrete choices of fluxes and by
determining
numerically the  local minima  of the full potential we confirm the
above behavior\footnote{In order to compare our result 
with \cite{Hebecker:2018yxs}, we define the warp factor in the infrared $w_{\rm
  IR}=e^A\sim ({\cal V}_w |Z|^2)^{1\over 6}$. Using the second
relation in \eqref{theomods}, one can write ${\cal V}_w\sim \tau^{3\over
  2}\sim (\log w_{\rm IR}^{-1})^{3\over 2}$, so that the mass of $Z$
can be expressed as
$m_Z^2\sim w_{\rm IR}^2/\log(w_{\rm IR}^{-1})^{3\over 2} M^2_{\rm pl}$. This 
agrees with the result in \cite{Hebecker:2018yxs}.}.

\subsection{AdS minimum}

\begin{figure}[ht]
  \centering
\hbox{  
\includegraphics[width=0.4\textwidth]{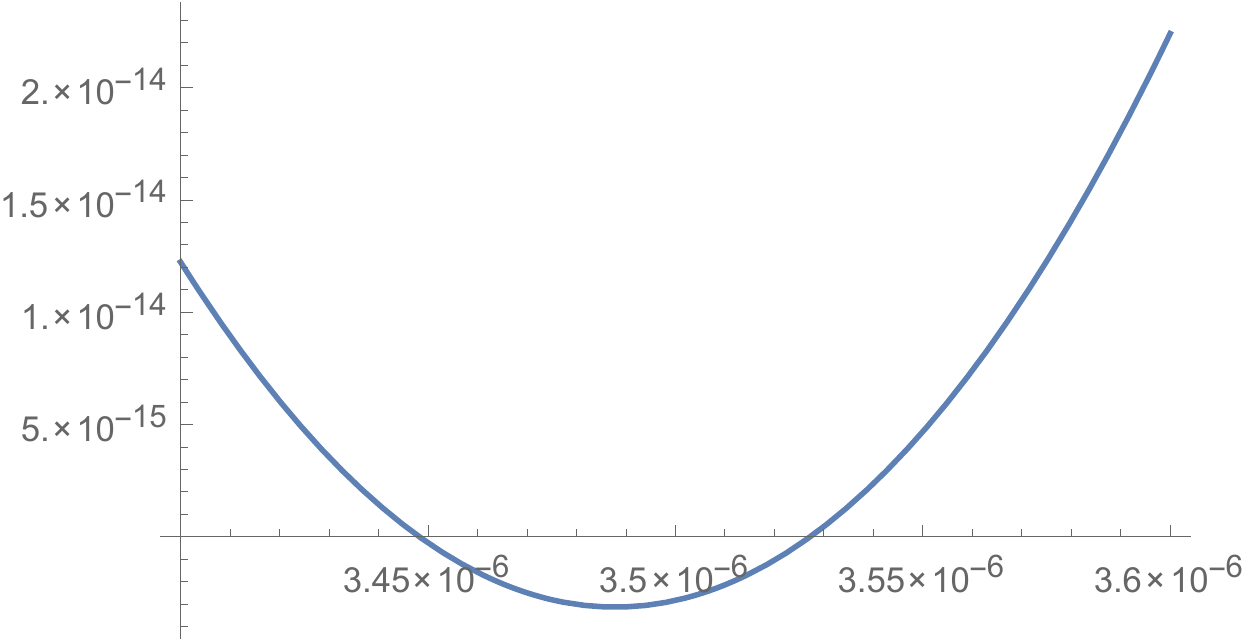}
\begin{picture}(0,0)
    \put(0,10){$\zeta$}
    \put(-148,90){$V$}
    \end{picture}
\hspace{1.0cm}
 \includegraphics[width=0.4\textwidth]{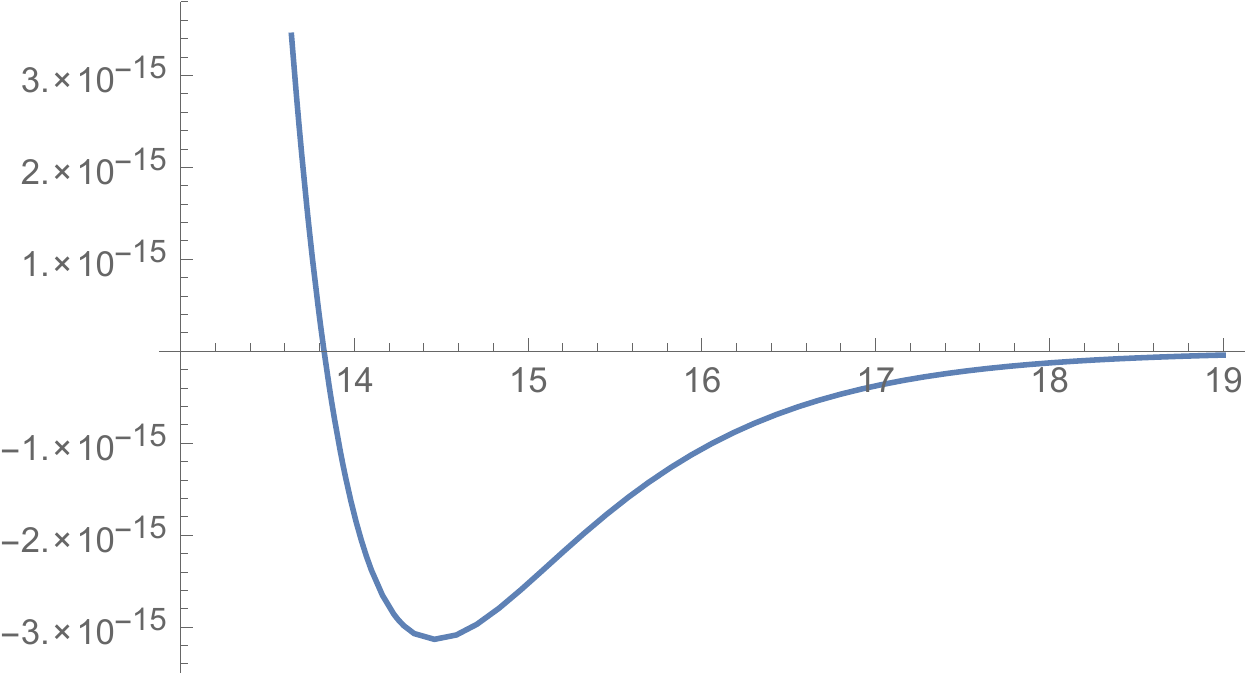}
\begin{picture}(0,0)
    \put(0,40){$\tau$}
    \put(-148,95){$V$}
    \end{picture}
}
  \caption{Left: $V(\zeta,\tau)$ for $\tau=14.47$,  Right:
     $V(\zeta,\tau)$ for $\zeta=3.49\cdot 10^{-6}$, both cases for $\sigma=0$, 
    $\theta=-\pi/2$ and the choice of
    parameters: $M=10$, $K=10$, $g_s=1/2$, $A=a=1$.}
  \label{fig:ads1}
\end{figure}
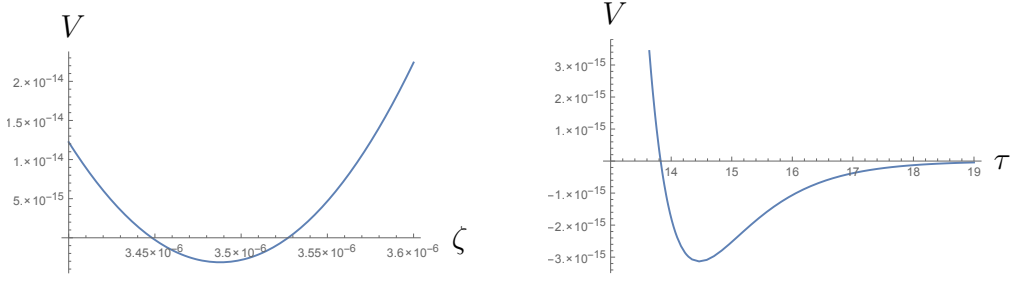

\begin{figure}[ht]
  \centering
  \includegraphics[width=0.65\textwidth]{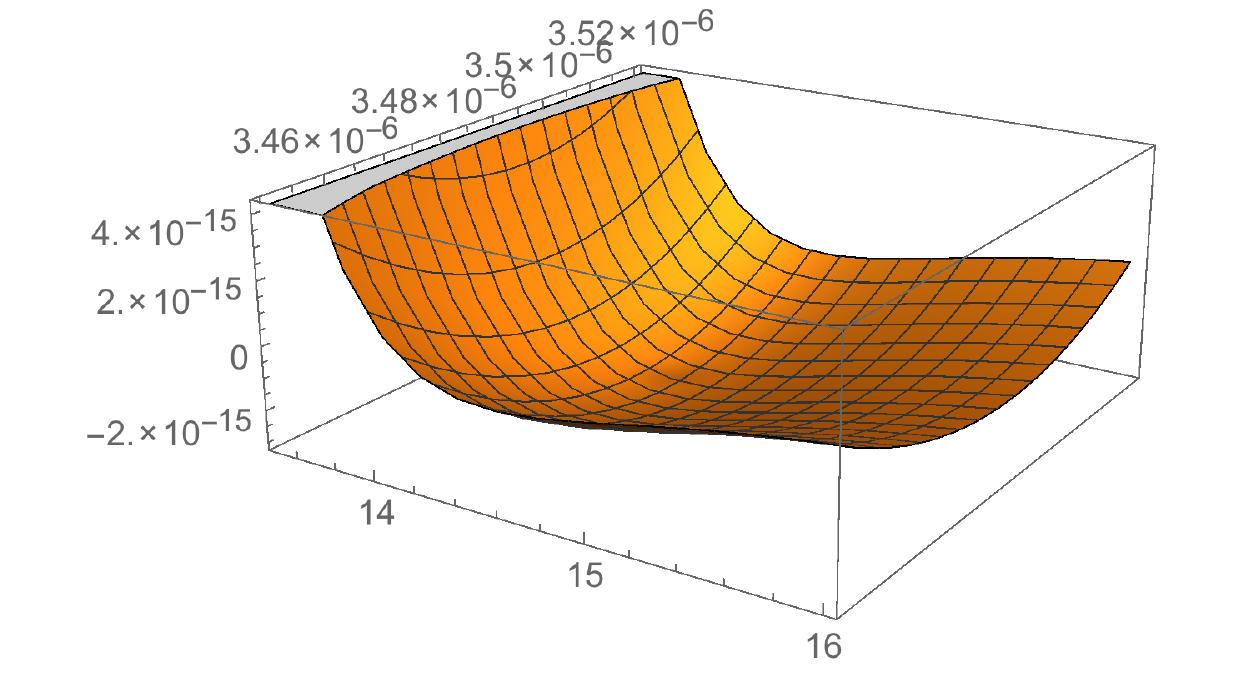}
\begin{picture}(0,0)
    \put(-80,8){$\tau$}
    \put(-117,143){$\zeta$}
    \end{picture}
  \caption{Potential $V(\zeta,\tau)$ for $\sigma=0$, $\theta=-\pi/2$ and the choice of
    parameters: $M=10$, $K=10$, $g_s=1/2$ and $A=a=1$.}
  \label{fig:ads2}
\end{figure}

Let us consider the full scalar potential without an  $\ov{\rm
  D3}$-brane. Then according to the previous  paragraph we 
expect to find a KKLT-like AdS$_4$ minimum of the full potential.
In the figures   \ref{fig:ads1} and \ref{fig:ads2} we
display the full potential in the region close to the minimum.
As can be seen, the behavior is consistent with our expectation.

\vspace{0.2cm}
%%%%%%%%%%%%
%%%%%%%%%%%%
%%%%%%%%%%%%

%%%%%%%%%%%%
%%%%%%%%%%%%

\noindent
The numerical AdS minimum lies at  $\zeta\vert_{\rm num}\approx 3.49\cdot 10^{-6}$, $\tau\vert_{\rm num} \approx 14.47$
with
\eq{
          V_0\vert_{\rm num}\approx -3.13\cdot 10^{-15} M_{\rm pl}^4\,.
}
Computing the mass eigenvalues for the saxions $\zeta$ and $\tau$ we
find
\eq{
                   m^2_\zeta\vert_{\rm num} \approx 9.38\cdot 10^{-7} M_{\rm pl}^2\,,\qquad
                   m^2_\tau\vert_{\rm num}\approx2.06\cdot 10^{-12}  M_{\rm pl}^2\,.
}
From the previous theoretical two step analysis, using
\eqref{theomods} we get for the values of the moduli in the AdS minimum  
\eq{   
\zeta|_{\rm theo}=3.49\cdot 10^{-6}\,,\qquad \tau\vert_{\rm theo}=14.47
}
which is in very remarkable agreement with the numerical one step result.
For the value of the potential in the minimum \eqref{AdScosmo} and for
the masses we find
\eq{
V_0\vert_{\rm theo}&\approx -1.27\cdot 10^{-15} M_{\rm pl}^4\,,\\
 m^2_\zeta\vert_{\rm theo} &\approx 3.18\cdot10^{-7}  M_{\rm pl}^2\,,\qquad
m^2_\tau\vert_{\rm theo}\approx 8.41\cdot 10^{-11} M_{\rm pl}^2
}
which are in the right ballpark. Therefore, we conclude that the true
AdS minimum of the full scalar potential is the one that we were able
to predict from the two-step procedure.

Thus, employing an effective low energy theory which takes warping
into account we seem to be  able to find a supersymmetric KKLT-like AdS
vacuum that by itself generates an exponentially small value of an
effective $W_0$. Of course one can add a sufficiently small value of
the superpotential $W_{\rm cs}$ without much changing our results. 
Let us make two remarks. First, the moduli satisfy ${\cal V}_w
|Z|^2\ll 1$ so that self-consistently they are fixed in the strongly
warped regime.
Second, as in KKLT, the AdS  vacuum satisfies scale separation, as
\eq{
           m_\tau^2\,  L^2_{\rm AdS}\sim -m_\tau^2 {M_{\rm pl}^2\over
             V_0}\sim (a \tau_0)^2 \gg 1
}
so that the vacuum can   indeed be considered four-dimensional.

\subsection{Uplift to de Sitter}

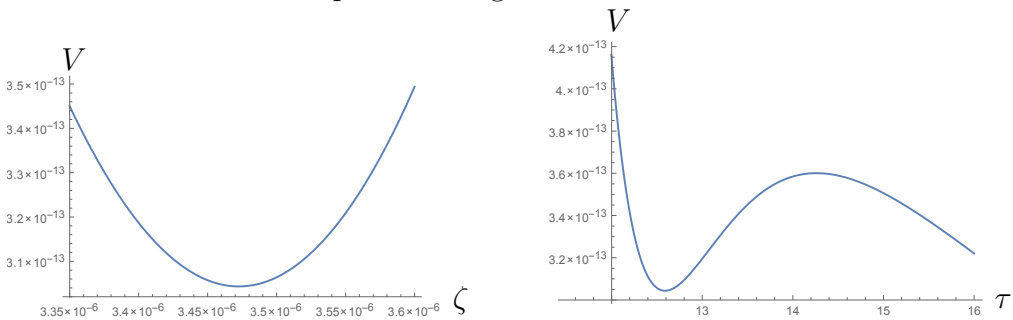
\begin{figure}[ht]
  \centering
\hbox{  
\includegraphics[width=0.4\textwidth]{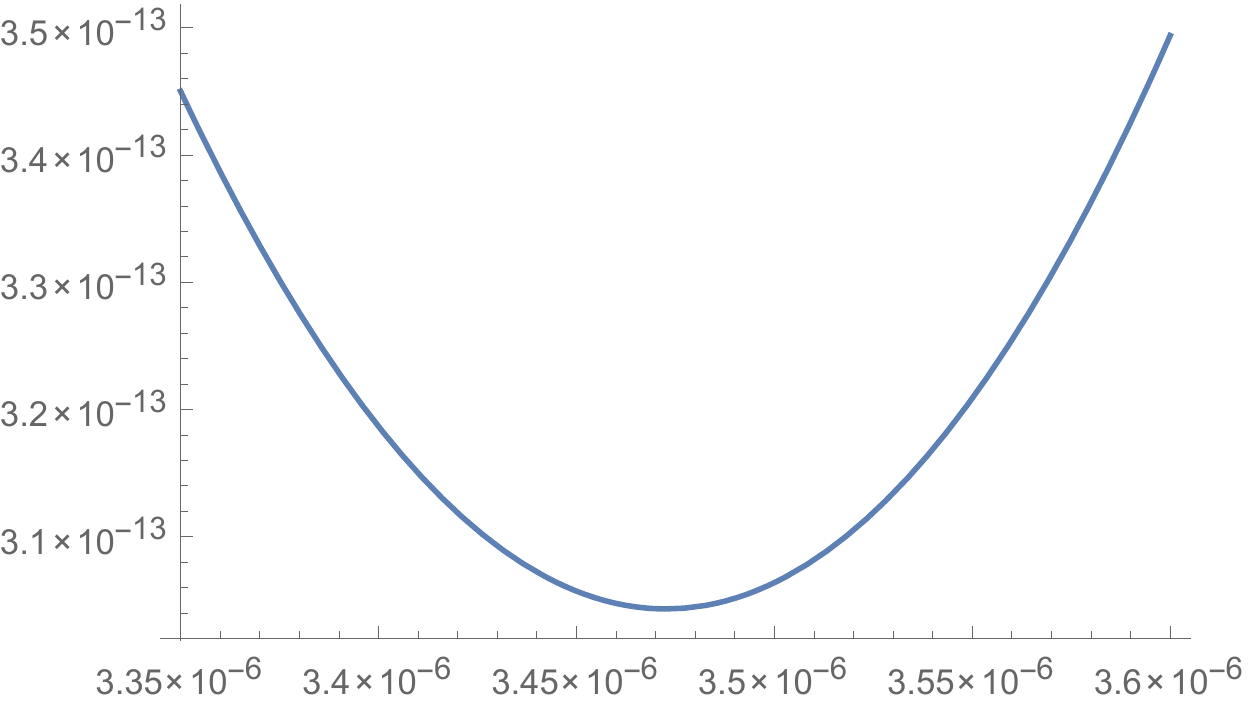}
\begin{picture}(0,0)
    \put(0,5){$\zeta$}
    \put(-148,95){$V$}
    \end{picture}
\hspace{1.0cm}
 \includegraphics[width=0.4\textwidth]{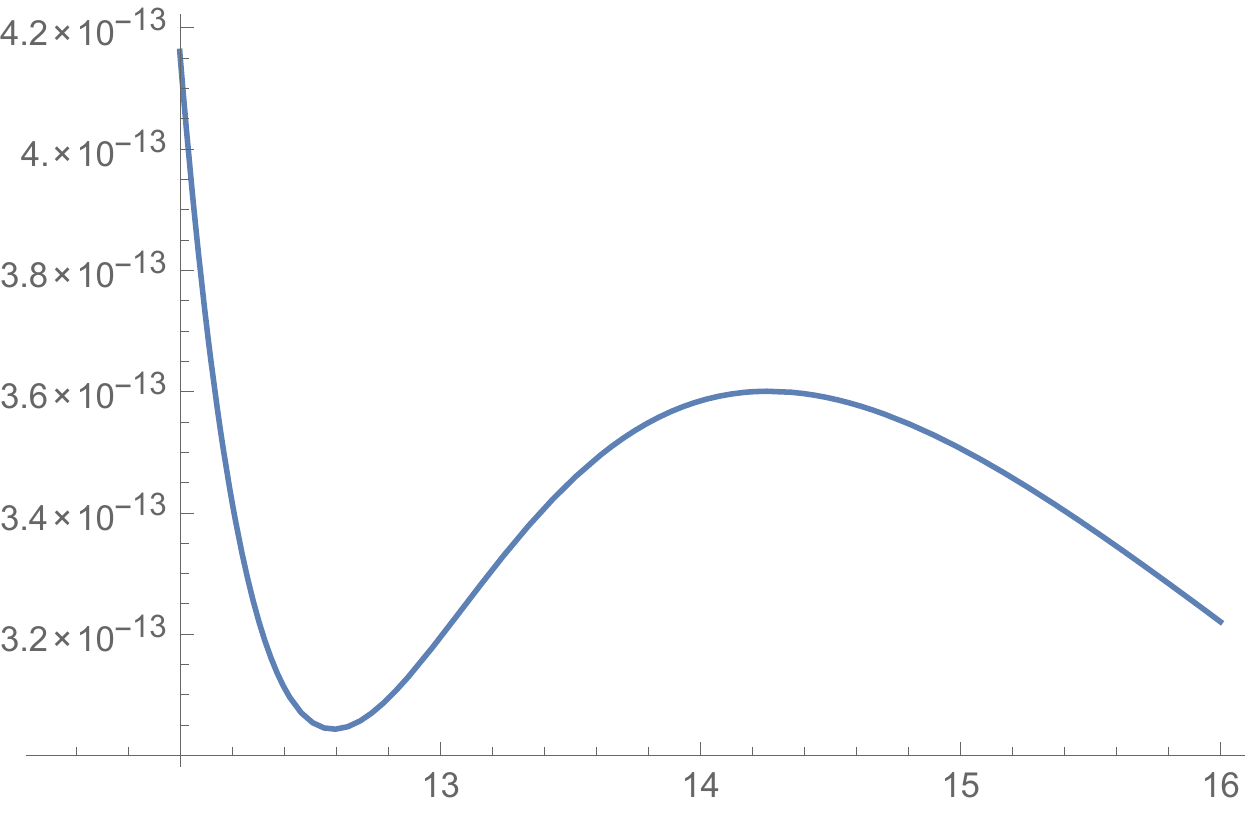}
\begin{picture}(0,0)
    \put(0,5){$\tau$}
    \put(-147,110){$V$}
    \end{picture}
}
 \caption{Left: $V(\zeta,\tau)$ for $\tau=12.59$,  Right:
     $V(\zeta,\tau)$ for $\zeta=3.47\cdot 10^{-6}$, both cases for $\sigma=0$, 
    $\theta=-\pi/2$ and the choice of
    parameters: $M=70$, $K=70$, $g_s=1/2$, $A=a=1$.}
  \label{fig:ds1}
\end{figure}

\begin{figure}[ht]
  \centering
  \includegraphics[width=0.7\textwidth]{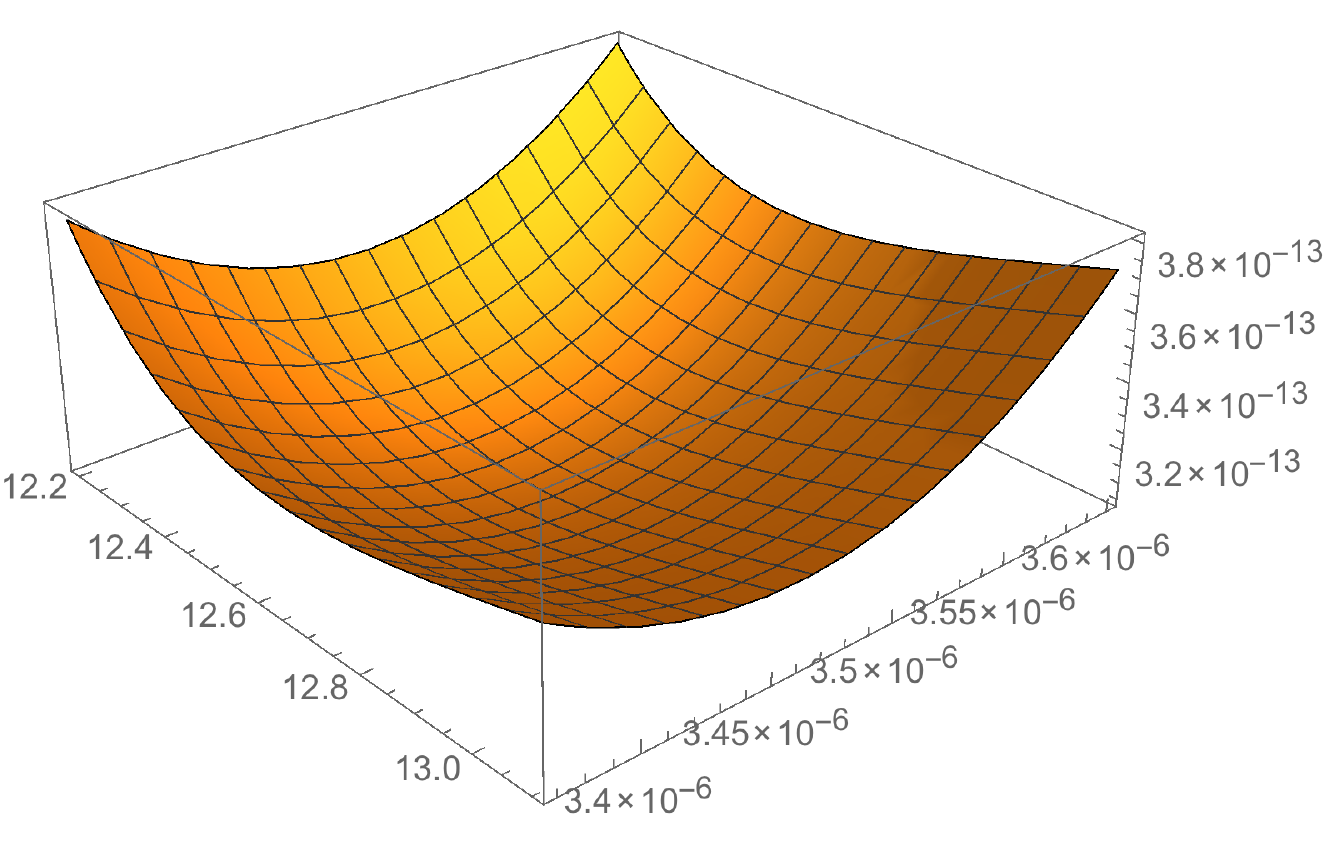}
\begin{picture}(0,0)
    \put(-37,68){$\zeta$}
    \put(-175,2){$\tau$}
    \end{picture}
 \caption{Potential $V(\zeta,\tau)$ for $\sigma=0$, $\theta=-\pi/2$ and the choice of
    parameters: $M=70$, $K=70$, $g_s=1/2$ and $A=a=1$.}
  \label{fig:ds2}
\end{figure}
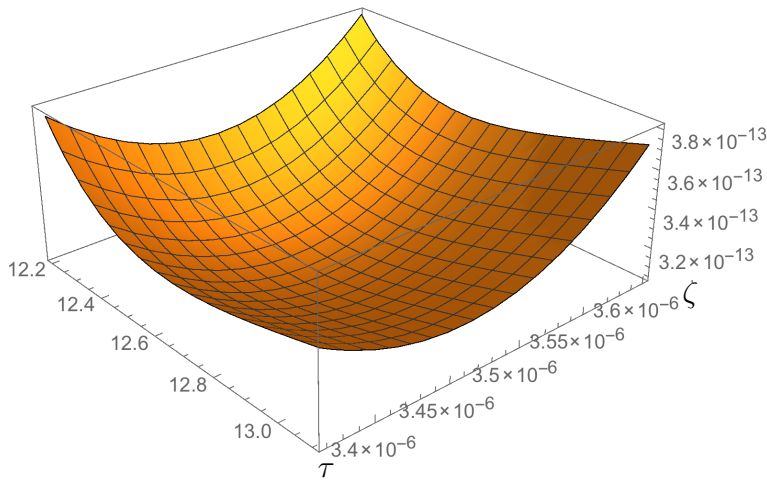

From here it is only one more step to uplift the AdS minimum to de
Sitter
by adding the contribution \eqref{Vantid3} of a $\ov{\rm  D3}$-brane
to the scalar potential.
Setting the AdS vacuum energy $V_0$ in \eqref{AdScosmo} equal to the energy of the 
$\ov{\rm  D3}$-brane \eqref{Vantid3} gives
\eq{
        {  |Z|^{2\over 3}\over {\cal V}^{4\over 3}}\sim {1\over (g_s M^2)^2} 
}
so that for an exponentially small $Z$ one expects large values of the flux  $M$.
As a proof of principle that such a metastable vacuum can indeed
exist, we provide a concrete numerical example in the figures
\ref{fig:ds1} and \ref{fig:ds2}.

%%%%%%%%%%%%
%%%%%%%%%%%%
%%%%%%%%%%%%

%%%%%%%%%%%%
%%%%%%%%%%%%
%%%%%%%%%%%%

Without assuming the existence of a ``tuning'' $0<|W_0|\ll 1$ 
(the choice $W_{\rm cs}=0$ is natural as it preserves supersymmetry)
in the string
landscape, we provided a dynamical KKLT scenario that seems to feature 
scale separated AdS minima and uplifted dS minima. For that purpose 
we employed an effective action for a conifold modulus and the overall K\"ahler 
modulus that seems to be valid and self-consistent in the strongly
warped regime. Therefore, either we have falsified the two
swampland conjectures forbidding such AdS and dS vacua or we have
overlooked an issue that spoils the validity of the effective
action used. 

It was discussed in the literature, whether the uplift by an
anti D3-brane in the warped throat might be too naive. If this was
correct then the dS vacua would be untrustworthy but the scale separated
AdS minima would survive. Could there be an issue that by one stroke
puts doubt to both types of minima?

\section{Swampland conjectures in the warped throat}
\label{sec_four}

In defining a Wilsonian effective action valid in the strongly warped regime,
we have assumed that all other massive states in the full string
theory are heavier than the conifold modulus (that by itself turned
out to be heavier than the K\"ahler modulus) of mass
\eq{
\label{massconiagain}
m_Z^2= {c\over g_s M^2}  \left( {|Z|\over {\cal V}_w}\right)^{\!{2\over  3}} M_{\rm pl}^2
} 
where $c$ is an order one coefficient.
It has been conjectured  that Wilsonian effective  field theories derived from
a UV complete theory of quantum gravity, have only a finite range of
validity. This behavior is described in the swampland distance
conjecture that says that by moving towards infinite distance in field
space, an infinite tower of states becomes exponentially light in the
proper field distance.
 
In contrast to e.g. the large complex structure point, the conifold point
is at finite proper field distance in the complex structure moduli space.
This is still true in the strongly warped case, as can be seen by
computing
\eq{
            \Phi=\int d\zeta \sqrt{G_{Z\ovmod Z}}\sim
            \left( {|Z| \over {\cal V}_w}\right)^{\! 1\over 3}
}
where the conifold point is at $Z=\Phi=0$. Therefore, one might hope that
even close to the conifold point the effective field theory is controllable.

However, it is well known  that Kaluza-Klein modes localized in the
strongly warped region might become dangerously light. 
These  KK modes in a warped throat have been investigated 
using various methods e.g. in \cite{Frey:2006wv, Burgess:2006mn,Shiu:2008ry,deAlwis:2016cty}.
Thus one might be  concerned  that  they become  parametrically lighter than the $Z$
modulus, thus spoiling the validity of the employed Wilsonian effective action in the
warped throat. 

In addition, as we have seen in section 2,   the
singularity in the complex structure moduli space is due to  integrating
out a non-perturbative state that becomes massless at the conifold
point \cite{Strominger:1995cz}.
Therefore, the question arises how the mass of this wrapped
$D3$-brane scales in the warped case.

\subsection{KK modes in the warped throat}
\label{sec_KK}

Let us first investigate the Kaluza-Klein modes.
 In order to be sure that we compare quantities
expressed in the same conventions, in this section we proceed by
computing the mass of the KK modes. We will employ two methods, 
the first will be a leading order approximation and the second a
numerical analysis.
Let us do the dimensional reduction of a ten-dimensional scalar field
$\Phi$  with mass $m$ to four  dimensions.
Starting with the  action 
\eq{
   S\sim \int d^{10}x \sqrt{-G} \Big(  G^{MN} \partial_M \Phi \partial_N \Phi
     + m^2 \Phi^2 \Big)\,
}
and making the usual warped ansatz for the ten-dimensional metric
\eq{   
          G=\left(\begin{matrix}   e^{2A(y)} g_4 & 0 \\ 0 &  e^{-2A(y)}
            \tilde g_{\rm CY} \end{matrix}\right)
}
the action can be written as
\eq{
                   S\sim\int d^4 x \sqrt{-g_4} \int d^6y \sqrt{\tilde g_{\rm
                       CY}}  \Big[ e^{-4A} g_4^{\mu\nu} &\partial_\mu
                     \Phi \partial_\nu \Phi +\\
                          & \tilde g_{\rm CY}^{mn} \partial_m
                     \Phi \partial_n \Phi + e^{-2A} m^2 \Phi^2 \Big]\,.
}
The resulting equation of motion for the field $\Phi(x,y)$ becomes
\eq{
               \Box_4\Phi + e^{4A} \tilde\nabla^m \tilde\nabla_m \Phi -e^{2A}m^2 \Phi=0\,.
}
Doing a product ansatz $\Phi(x,y)=\varphi(x) \chi(y)$ the
four-dimensional KK masses $m_{\rm KK}^2$ are given by the eigenvalues of the
six-dimensional warped Laplace equation
\eq{
\label{laplacewarp}
        e^{4A(y)} \tilde\nabla^m \tilde\nabla_m \chi(y) - m^2\, e^{2A(y)}
        \chi(y)=-m^2_{\rm KK}\, \chi(y)\,.
}
We are heading for the lightest modes, which are expected to arise
from the KK modes 
 of the four-dimensional components of the metric
$g_{4\,\mu\nu}(x,y)$. The zero mode is the 4D graviton that is the
lowest excitation of the closed string. Placing such a closed string
deep into the throat region, we expect to find highly red-shifted KK masses.
Therefore, we set  $m=0$ and note that at linear order  KK modes of the 4D
metric  are also governed by the same Laplace equation (see \cite{Shiu:2008ry})
\eq{
\label{laplacetos}
        e^{4A(y)}\, \widetilde\nabla^2_{\rm CY} \,\chi(y) =-m^2_{\rm KK}\, \chi(y)\,.
}
Actually, one now has to solve this equation on the entire Calabi-Yau
manifold for a point in complex structure moduli space that is very
close to a conifold singularity. This is a horrendous task that is
beyond the scope of this paper.

Here, we take a simpler approach and first look for local solutions 
that are supported close to the
tip of the cone of the KS solution \eqref{KSmetric}. These are the
ones which are expected to yield small red-shifted masses.
For this purpose, we take the local CY metric of the KS throat and 
evaluate the Laplacian $\widetilde\nabla^2_{\rm CY}$ for solutions
that do only depend on the radial direction $y$ and are constant
on the $S^2\times S^3$ base of the cone. These are expected 
to be the ones that have minimal mass. We compute the relevant Laplace equation from the KS metric
\eqref{KSmetric} and warp factor \eqref{warpKS}
\eq{
\label{fulldgl}
               {2^{1/3}} {({\cal V} |Z|^2)^{1\over
                   3}\over g_s^{3/2}  M^2} {1\over {\cal I}(y)}\Big[ 3 K^2(y)\, \partial_y^2 \chi(y)
               +{4} {\partial_y \chi(y)\over \sinh(y)\, K(y)}
                 \Big]=-\alpha' m^2_{\rm KK} \,\chi(y)\,.
}

\subsubsection*{Leading order approximation}

Before we solve this differential equation numerically, to get an idea
what the solutions might look like, we
expand all  quantities up to leading order around $y=0$. Using 
\eq{
       K(y)&=\left({2\over 3}\right)^{1\over 3} 
+ O(y^2)\\
      \sinh(y)\, K(y)&= \left({2\over 3}\right)^{1\over 3} y + O(y^3)\\[0.1cm]
    {\cal I}(y)&=\kappa 
+ O(y^2)
}
with $\kappa\approx 0.72$, we arrive at 
\eq{
               {2\cdot 3^{1/3} \over \kappa} {({\cal V} |Z|^2)^{1\over
                   3}\over g_s^{3/2} M^2}\Big[ \partial_y^2 \chi
               +{2\over
                   y} \partial_y \chi\Big]=-\alpha' m^2_{\rm KK} \,\chi\,.
}
Up to some scaling factors, this is the spherical Bessel  differential equation
\eq{
           \Big[ \partial_y^2 \chi
               +{2\over
                   y} \partial_y \chi+k^2 \chi\Big]=0
}      
whose solution with Neumann boundary conditions at $y=0$ is
\eq{
             \chi(y) = {\sin(ky)\over ky}             
}
where $k_n= f_n/y_{\rm UV}$ is expected to be quantized by 
imposing (Neumann) boundary
conditions at the UV end $y_{\rm UV}$ of the throat.
Here $f_n$ denotes the solutions of the equation $\tan f=f$ that are
approximately $f_n\approx (2n+1)\pi/2$ with $n\ge 1$.

Therefore, the KK masses of these localized solution scale 
\eq{
\label{KKmassesth}
        m^2_{\rm KK}={2\cdot 3^{1/3} \over \kappa}     {f_n^2\, ({\cal V}_w |Z|^2)^{1\over
                   3}\over g_s^{3/2} (M y_{\rm UV})^2}\, M_{\rm s}^2\,.
}
Note that, with respect to $M_s$, $g_s$,  $M$ and $({\cal V}_w |Z|^2)$
this scales precisely in the same way as the mass \eqref{massconiagain} of the
conifold modulus, so that
\eq{
                   {m^2_{\rm KK}\over m_Z^2} = c \, {f_n^2\over   y_{\rm UV}^2}
}
where $c$ is an order one coefficient.
Note that $y_{\rm UV}$ is bounded from below by
  \eqref{lowboundy} that is  weaker than imposing $y_{\rm UV}>1$. In the latter
  regime there are finitely many KK
modes that have a mass lighter than the conifold modulus. This
indicates that the employed effective action might be at the edge of
reliability. We will further analyze this important question in the
upcoming sections.

In Einstein frame, massive bulk string excitations have a mass 
$m^2_{\rm  str}\sim g_s^{1/2} M^2_s$.
If placed into the warped throat we have checked that there exist 
localized solutions of \eqref{laplacewarp} leading to KK masses that are shifted
up by
\eq{
\label{redshiftstring}
        m^2_{\rm str, throat}\sim {1\over M} ({\cal V}_w |Z|^2)^{1\over
          3} M_s^2\,.
}
With respect to ${\cal V}_w$ and $Z$ this scales in the same way as
the mass of the conifold modulus and the KK modes. Note, that in this
way each of
the string modes comes with a whole tower of KK excitations with spacing
of the order \eqref{KKmassesth}.

\subsubsection*{Numerical solution of warped Laplace equation}

\begin{figure}[ht]
  \centering
\hbox{  \includegraphics[width=0.45\textwidth]{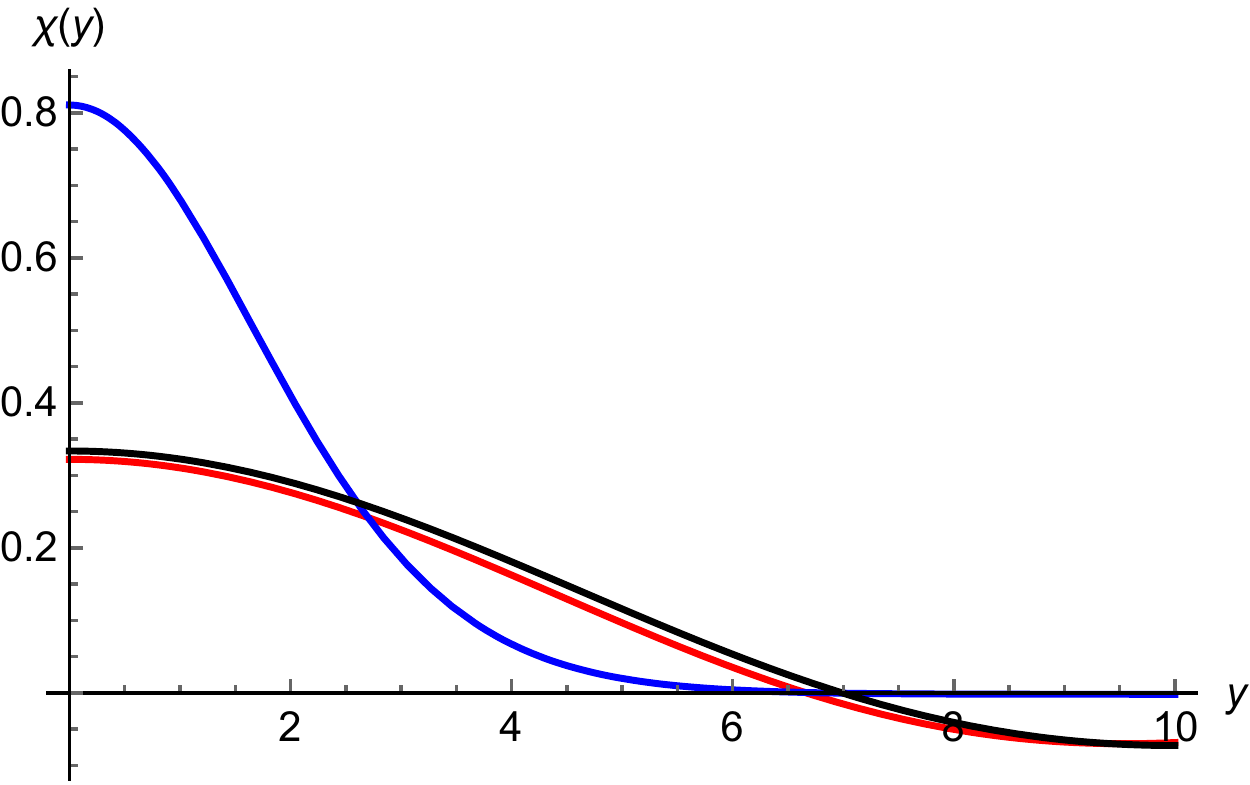}
   \hspace{0.7cm} \includegraphics[width=0.45\textwidth]{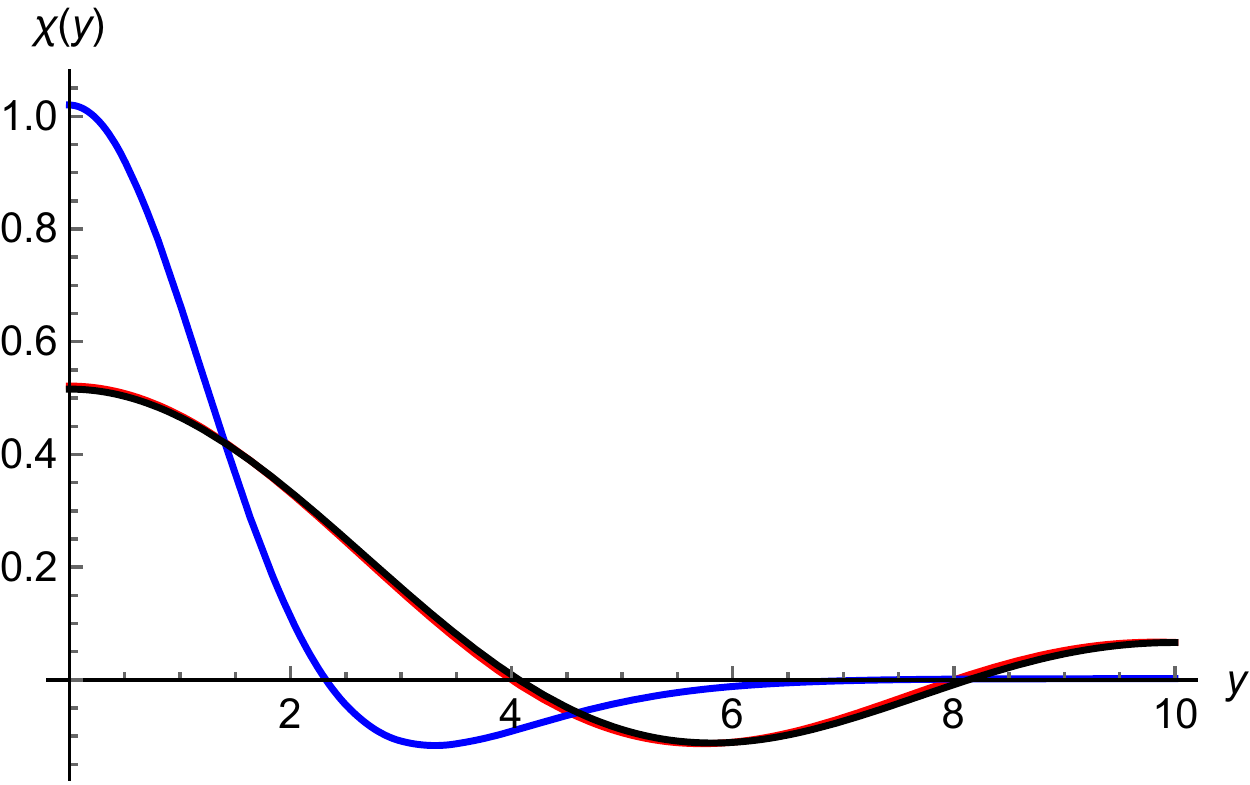}}
  \caption{Eigenfunction of the numerical solution (blue) and analytical
    approximation (red) of the first (left) and second (right) radial mode.}
  \label{wave1}
\end{figure}

For $y>1$ we do not expect our leading order approximation to be valid
so that a full numerical analysis of the solution of \eqref{fulldgl} is
necessary. Except for the radial dependence of ${\cal I}(y)$, this is a one dimensional differential equation with Neumann boundary conditions. This function could only be evaluated numerically. To obtain an expression which can be inserted into the numerical procedure, the function was sampled at $5000$ points in the interval $[0,50]$ and interpolated using a degree three polynomial.
Figure \ref{wave1} shows the first and second
eigenfunctions of the approximate analytical solution as well as the
numerical solution. 

\noindent
The functions are normalized such that the integral over the absolute
values squared is equal to one. Only the radial contribution is shown, i.e. all prefactors are set to one:
\begin{equation}
{({\cal V}_w |Z|^2)^{1\over
                   3}\over g_s M^2\alpha'}=1\,.
\end{equation}

To estimate the numerical errors made in solving the differential
equations, the same methods were used to solve the spherical Bessel
equation numerically, the results are shown in all figures in
black. The functions as well as eigenvalues agree with the analytical
result, showing that the numerical errors are small. We notice that the numerical
functions are shifted towards small $y$ relative to the analytical
spherical Bessel functions, improving the localization in the warped 
throat.

\begin{figure}[ht]
  \centering
 \hbox{ \includegraphics[width=0.45\textwidth]{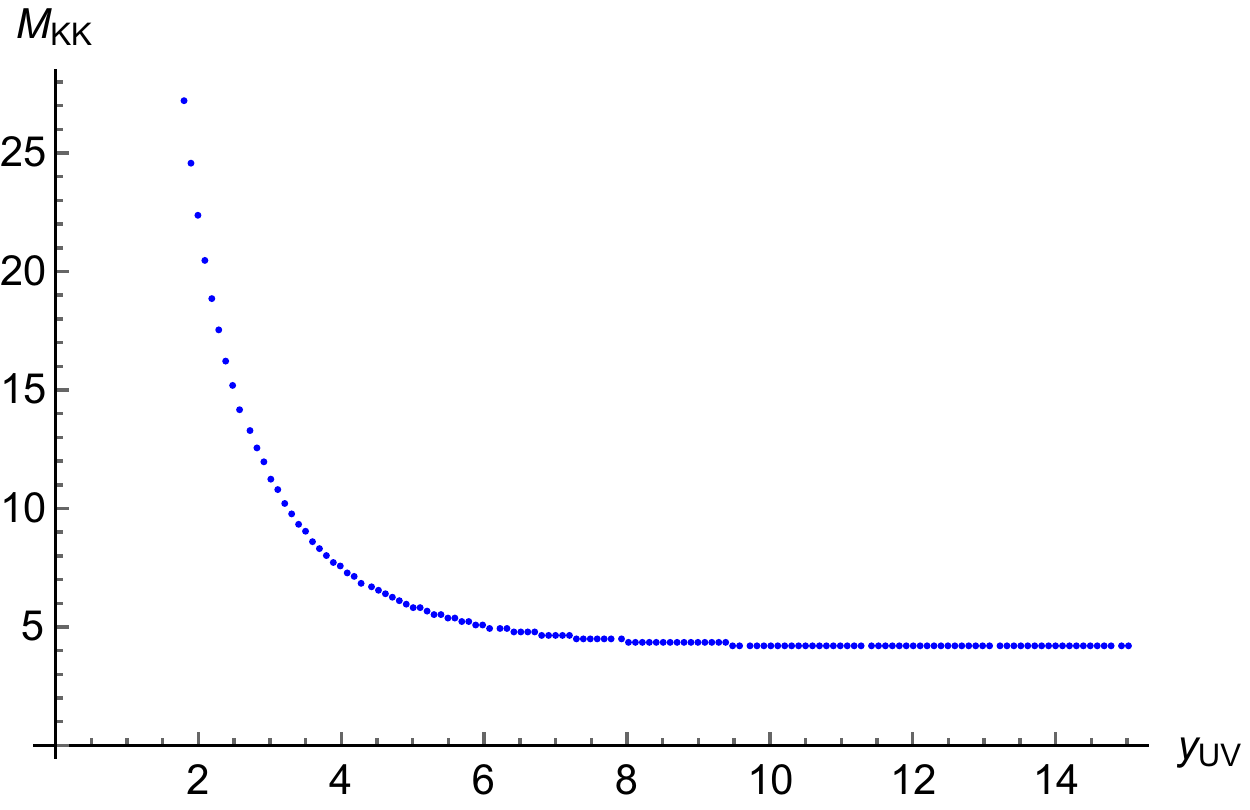}
\hspace{0.4cm}
 \includegraphics[width=0.45\textwidth]{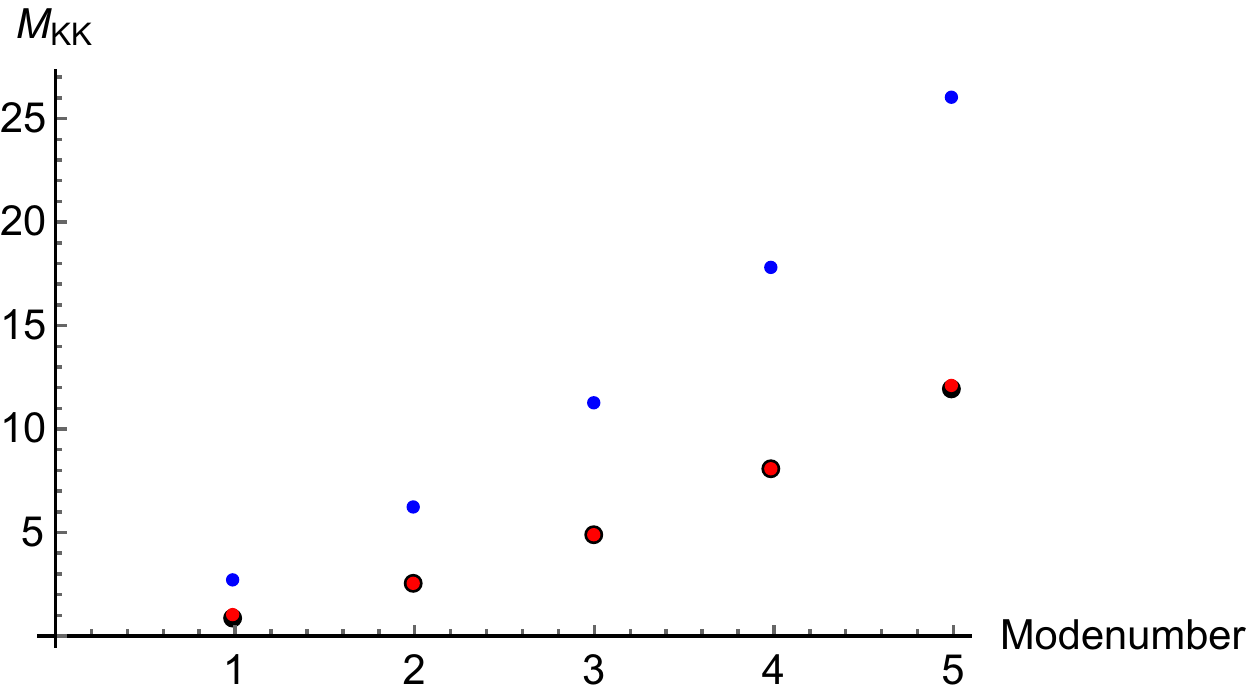}
}
\caption{Left: The first eigenvalue of the numerical solution for
    different $y_{UV}$. Right: Eigenvalues of the numerical solution (blue) and analytical approximation (red) of the first five radial modes. The numerical evaluation of the spherical Bessel equation is shown in black.}
  \label{radialDependence}
\end{figure}

The eigenvalues of the numerical solution scale approximately like
$1/y_{UV}$ for small values of $y_{UV}$ and approach an asymptotic
value for  $y_{UV} \gtrapprox 10$ due to the localization of the
functions at small $y$. The left hand side of figure  \ref{radialDependence} shows this
behavior exemplary for the case of the  first eigenmode.
The right figure shows the mass eigenvalues obtained via   the analytical and the
numerical method.

\newpage

\subsection{The swampland distance conjecture}
\label{sec_4.2}

In this subsection, we establish contact of the KK spectrum with the
recent discussions of the  swampland distance conjecture (SDC), in
particular its emergent nature. First, let us recall the latter point 
in the large distance regime.

\subsubsection*{The emergence of the swampland distance conjecture}

The behavior of a tower of modes becoming lighter than the cut-off has recently been claimed
\cite{Heidenreich:2017sim,Grimm:2018ohb,Heidenreich:2018kpg,Corvilain:2018lgw}
to underly the
swampland distance conjecture of quantum gravity. Closely following
\cite{Corvilain:2018lgw}, let us briefly repeat how the logic
goes.  Say one has a light field $\phi$ and a tower of massive states
$h_n$ governed by an effective action
\eq{
\label{towereffact}
           S=M_{\rm pl}^2 \int d^4 x \left( {1\over 2} g_{\phi\phi} \partial_\mu\phi  \partial^\mu
           \phi+   \sum_n  {1\over 2}  \partial_\mu h_n \partial^\mu
           h_n +  {1\over 2} m^2_n(\phi)  h_n^2 \right)\,
}
where the fields are dimensionless.
The masses of the tower are assumed to be discretized as $m_n=n \Delta m(\phi)$ and
to depend on the value of the field $\phi$.  The question is what
happens if some of these states become lighter than the natural
cut-off scale, which for a theory of quantum  gravity is the so-called
species scale\footnote{We notice that at this scale the one-loop correction to the Planck-scale
\begin{equation*}
                 M^2_{\rm pl}(\mu) = M^2_{\rm pl}(0) - {\mu^2\over
                   12\pi} N_{\rm sp}
\end{equation*} 
from integrating out the $N_{\rm sp}$ species becomes of the order of the IR
value  $M_{\rm pl}(0)$.} (see for example \cite{ArkaniHamed:2005yv,Distler:2005hi,Dimopoulos:2005ac,Dvali:2007wp,Dvali:2007hz})

\eq{
                     \tilde \Lambda_{\rm sp}={M_{\rm pl}\over 
                     \sqrt{N_{\rm sp}}}
}
where $N_{\rm sp}$ denotes the number of states/species with a mass lower than the cut-off,
i.e. $N_{\rm sp}=\tilde \Lambda_{\rm sp}/\Delta m(\phi)$. The last two
relations can be solved as
\eq{
              \tilde \Lambda_{\rm sp}=\Big( M^2_{\rm pl}\,\Delta
                m(\phi) \Big)^{1\over 3}\,,\qquad
              N_{\rm sp}=\left( {M_{\rm pl}\over \Delta
                m(\phi) }\right)^{2\over 3}\,.
}
Clearly, those states which are lighter than $ \tilde \Lambda_{\rm sp}$ spoil a
Wilsonian effective action just for the field $\phi$ below the
species bound. Including them in the effective action (as in
\eqref{towereffact}), one can compute their contribution to the
one-loop 1PI effective action for the field $\phi$. 
For bosonic states, as shown in \cite{Grimm:2018ohb}, their effect on the field space metric
$g_{\phi\phi}$ is
\eq{
                  g_{\phi\phi}^{\rm 1-loop}&\sim M_{\rm pl}^{-2}\sum_{n=1}^{N_{\rm sp}}
                  \Big( \partial_\phi m_n(\phi) \Big)^2
                  =M_{\rm pl}^{-2}\Big( \partial_\phi \Delta m(\phi) \Big)^2\sum_{n=1}^{N_{\rm sp}}
                  n^2\\
                     &=  {N_{\rm sp}^3\over M_{\rm pl}^2}  \Big( \partial_\phi \Delta
                     m(\phi) \Big)^2
                   = \left({\partial_\phi \Delta
                     m(\phi)\over \Delta
                     m(\phi)}\right)^2\,.
}
For fermions, as in \eqref{eq:IntegrateOutBoseFermi}, the relevant UV cutoff is the associated species scale $\Lambda_{\rm UV}=\tilde{\Lambda}_{\rm sp}$ and we compute
\eq{
                  g_{\phi\phi}^{\rm 1-loop}&\sim M_{\rm pl}^{-2}\sum_{n=1}^{N_{\rm sp}}
                  \Big( \partial_\phi m_n(\phi) \Big)^2\log\left(\frac{\tilde{\Lambda}_{\rm sp}^2}{m_n(\phi)^2}\right)\\
                  &=M_{\rm pl}^{-2}\Big( \partial_\phi \Delta m(\phi) \Big)^2\sum_{n=1}^{N_{\rm sp}}
                  n^2 \log\left(\frac{N_{\rm sp}^2}{n^2}\right)\\
                     &\leq \frac23 {N_{\rm sp}^3\over M_{\rm pl}^2}  \Big( \partial_\phi \Delta
                     m(\phi) \Big)^2
                   = \frac23 \left({\partial_\phi \Delta
                     m(\phi)\over \Delta
                     m(\phi)}\right)^2\,.
}
Here we have used an integral approximation of the sum. As $N_{\rm
  sp}\to \infty$ the inequality becomes saturated. In any case we see
that for such an evenly spaced tower of states, the contribution from
fermions is always of the same functional form as for
bosons\footnote{Note that this is in contrast to the case of
  integrating out the single wrapped D3-brane \eqref{eq:IntegrateOutD3Unwarped}, where the fermion contribution was dominant because of the large logarithm.}. 
For this reason we will restrict our attention to bosonic states in the 
following section.

In both cases, the proper field distance can be evaluated as
\eq{
            d(\phi_0,\phi_1)\sim \int_{\phi_0}^{\phi_1} d\phi\,
            \sqrt{g_{\phi\phi}} \sim \log\left({\Delta
                     m(\phi_1)\over \Delta
                     m(\phi_0)}\right)
}
showing the typical logarithmic behavior.  Therefore, one can write
\eq{
          \Delta m(\phi_0)\sim \Delta
                     m(\phi_1) \, e^{-\gamma d(\phi_0,\phi_1)}
}
so that at infinite distance in field space, a tower of states becomes
exponentially light. In addition the number of light species also
increases exponentially with the proper field distance.

This is the swampland distance conjecture, which was
claimed to be satisfied for every point of infinite distance in field
space. It has been emphasized in \cite{Grimm:2018ohb} that this IR property is
emergent, in the sense that  it follows from integrating out UV  states of mass below the
species bound.

\subsubsection*{Integrating out KK modes in the warped throat}

Now, we would like to apply the same logic to the region close to the
conifold locus. Since we do not yet know the cut-off of the effective
action in the warped throat, we require that the metric on moduli
space is emerging in the same way as above by integrating out light  bosonic
and fermionic  modes.

In the regime $g_s M^2\gg 1$ the tower of KK modes with spacing
\eq{
   \Delta m \approx {1\over \sqrt{g_s M^2} \, y_{\rm UV}} \left({|Z|\over {\cal
          V}_w}\right)^{\!{1\over 3}} M_{\rm pl}
}
is supposed to be lighter than the cut-off. Integrating out these
light gravitationally coupled KK modes leads to a one-loop correction
to the field space metric
\begin{equation}
\label{eq:KK1loop}
	\begin{aligned}
		 g_{Z\ovmod Z}^{\rm 1-loop}&\sim M_{\rm pl}^{-2}\sum_{n=1}^{N_{\rm sp}}
                  \Big( \partial_Z m_n(Z) \Big)^2\sim \sum_{n=1}^{N_{\rm sp}}n^2 \left(\frac{1}{\sqrt{g_s M^2}y_{\rm UV}}\frac{1}{(\mathcal{V}_w|Z|)^{1/3}}\right)^2\\[0.2cm]
                  &\sim N_{\rm sp}^3\frac{1}{g_s M^2y_{\rm UV}^2}\frac{1}{(\mathcal{V}_w|Z|)^{2/3}}\;.
	\end{aligned}
\end{equation}
Consistency with the picture of kinetic terms arising from integrating out fields in the UV demands that the parametric scaling of this contribution matches the tree level result
\begin{equation}
\label{hannover96}
	g_{Z\ovmod Z}^{\rm 1-loop}\sim \frac{g_sM^2}{(\mathcal{V}_w |Z|^2)^{2/3}}\;.
\end{equation}
Enforcing this scaling in \eqref{eq:KK1loop} constrains the number of
light KK species in the effective description to scale as
\begin{equation}
	N_{\rm sp}\sim\left(g_s M^2 y_{\rm UV}\right)^{2/3}\;.
\end{equation}
Note that due to the lower bound \eqref{lowboundy}, this number is
guaranteed to satisfy $N_{\rm sp}\gtrsim |M|^{2/3}$. Thus, there is a
finite number of KK modes whose mass is lighter than the species scale.

Using this scaling, analogously one finds that the corrections $g_{Z\ovmod[1] T}^{\rm 1-loop}$ and $g_{T\ovmod[1] T}^{\rm 1-loop}$ are proportional to the tree-level expressions following from the K\"ahler potential $K\sim g_s M^2 |Z|^{2\over 3}/(T+\ovmod[1] T)$.
As a result, a consistent effective description of the warped throat
should accommodate at most $N_{\rm sp}$ light KK modes and thus should have a cutoff of at most
\begin{equation}
	\tilde\Lambda_{\rm sp}\sim N_{\rm sp}\Delta m \sim \left({ g_s M^2\over y_{\rm
              UV}^2}\right)^{\!{1\over 6}} \left({|Z|\over {\cal
          V}_w}\right)^{\!{1\over 3}} M_{\rm pl}\;.
\end{equation}
In analogy to the ``gravitational'' species scale $\Lambda_{\rm
  sp}=M_{\rm pl}/\sqrt{N_{\rm sp}}$ we can interpret this scale as a
generalized species 
scale\footnote{At this scale the one-loop correction to the Planck-scale
$M^2_{\rm pl}(\mu) = M^2_{\rm pl}(0) - {\mu^2\over 12\pi} N_{\rm sp}$ 
becomes of the order of the cut-off scale $\Lambda$. }
\eq{
\tilde{\Lambda}_{\rm sp}=\frac{\Lambda}{\sqrt{N_{\rm sp}}}
}
for an effective gravity theory with a cut-off 
\eq{
    \Lambda\sim \sqrt{g_s M^2} \left({|Z|\over {\cal
          V}_w}\right)^{\!{1\over 3}} M_{\rm pl}\,.
}

\begin{figure}[ht]
%\centering
\vspace{-2ex}
  \begin{tikzpicture}
    \node[anchor=south west,inner sep=0] (modulispace) at (0,0) {\includegraphics[width=0.8\textwidth]{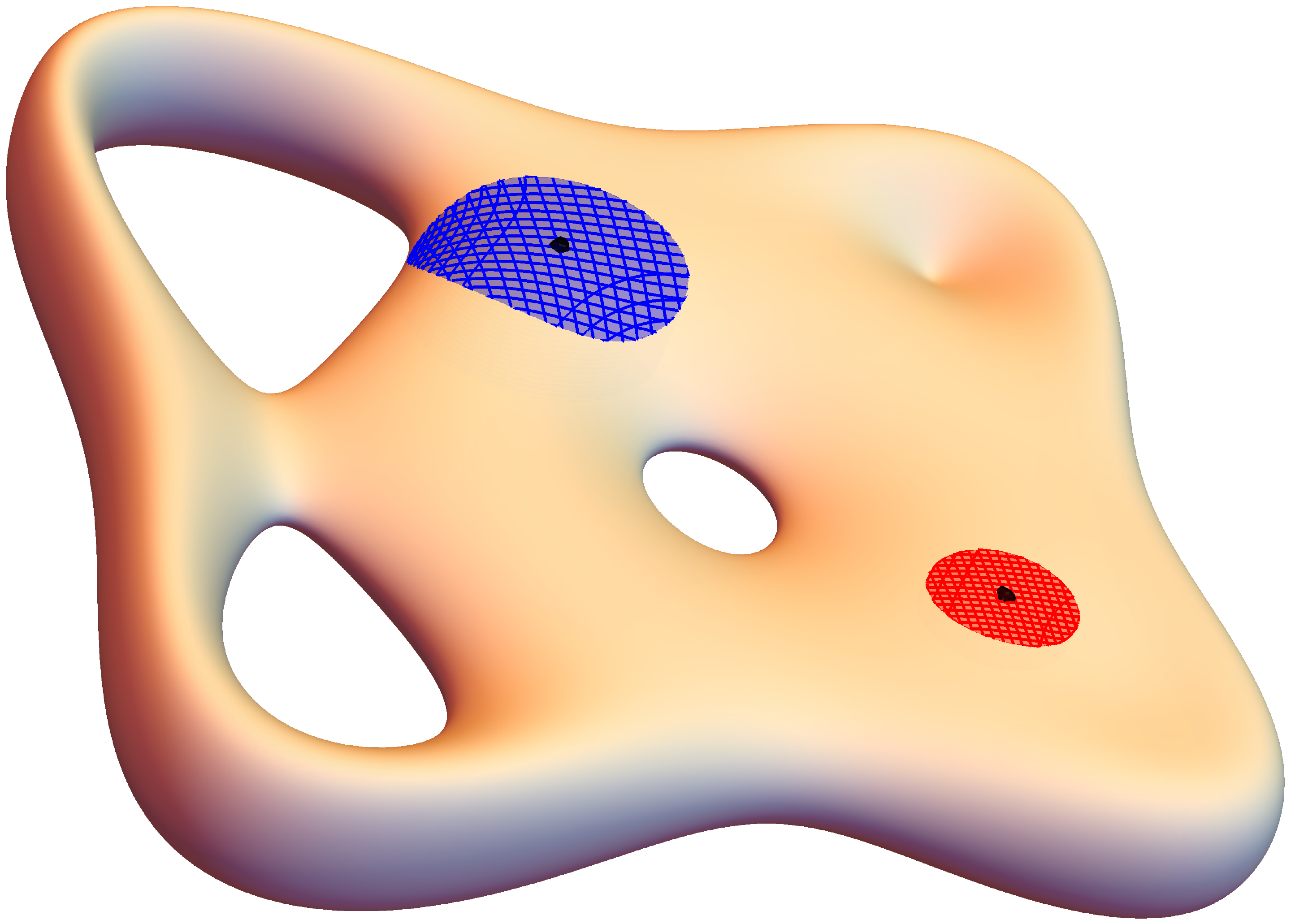}};
    %\draw[help lines,xstep=1,ystep=1] (0,0) grid (14,9);
    %\foreach \x in {0,1,...,13} { \node [anchor=north] at (\x,0) {\x}; }
    %\foreach \y in {0,1,...,8} { \node [anchor=east] at (0,\y) {\y}; }
    \node (LCS) at (12.1,4.1) {$\begin{aligned}\Delta m&\sim \exp(-\alpha\Phi) M_\text{pl}\\
                                          N&\sim \exp\left(\tfrac{2}{3}\alpha\Phi\right)\\
                                          \Lambda&\sim M_\text{pl}
                           \end{aligned}$};
    \draw (9.03,2.97) -- (10.45,4.13);
    \node (Conifold) at (6.1,8.15) {$\begin{aligned}\Delta m&\sim \Phi M_\text{pl}\\
                                               N&\sim \text{const.}\\
                                               \Lambda&\sim \Phi M_\text{pl}
                                \end{aligned}$};
    \draw (5.04,6.11) -- (6,7.3);
\node (KKLT) at (3.73,8.79) {$\begin{aligned}{\rm KKLT}
                                \end{aligned}$};
  \draw (3.92,8.61) -- (4.5,6.11);

\node (trans) at (12.40,1.70) {trans-Planckian};
  \draw (9.25,2.71) -- (11.95,1.83);
  \end{tikzpicture}
  \caption{A sketch of the complex structure ``moduli space'' of a
    warped Calabi-Yau. The red/blue regions are a neighborhood of the
    large complex structure and conifold points where towers of  modes become
    lighter than the cut-off scale.} 
  \label{fig:ModuliSpace}
\end{figure}
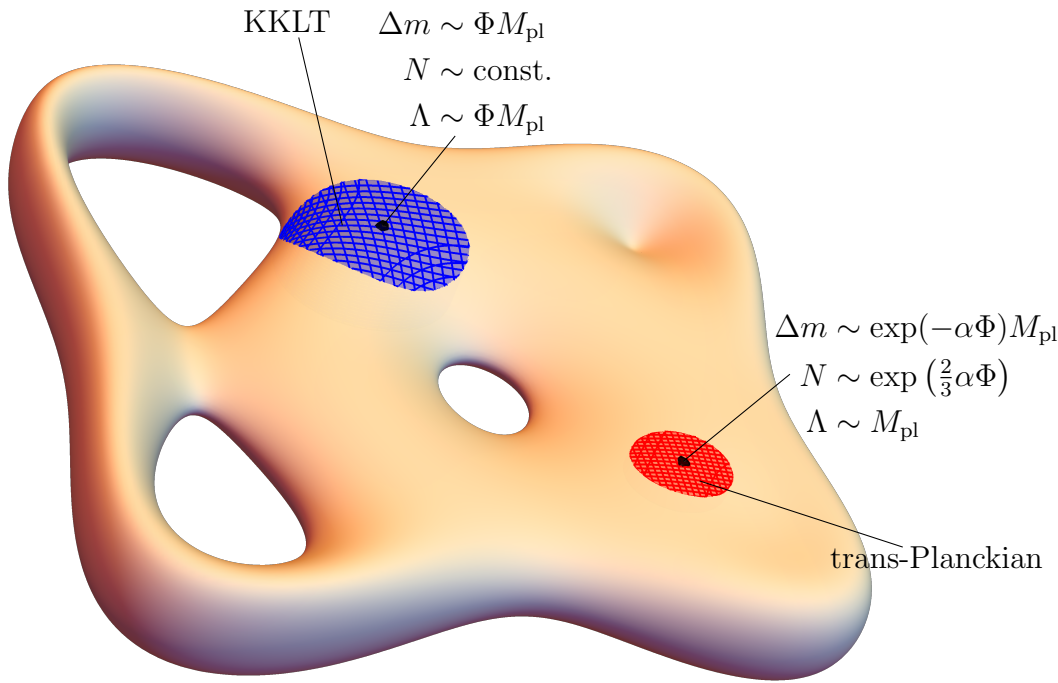

In contrast to the emergence of the SDC at large volume, here the ultimate
cut-off $\Lambda$ is also field dependent. This  implies  a finite
distance of the conifold point in complex structure moduli space
\eq{
      \Phi=d(0,|Z_0|)\sim \int_0^{|Z_0|} \sqrt{ g_{Z \ovmod Z}}\sim \sqrt{g_s
        M^2} \left({|Z_0|\over {\cal
          V}_w}\right)^{\!{1\over 3}}\sim {\Lambda\over M_{\rm pl}} \,,
}
where $\Phi<1$ is the canonically normalized field corresponding to $Z$.
In terms of $\Phi$ the relevant quantities become
\eq{
          \Lambda\sim \Phi\, M_{\rm pl}\,,\qquad
           \Delta m\sim {\Phi\over g_s M^2\, y_{\rm UV}} \, M_{\rm
             pl}\,,\qquad
            \tilde\Lambda\sim {\Phi\,\over(  g_s M^2\, y_{\rm
                UV})^{1\over 3}} \, M_{\rm pl}
}
with still $N_{\rm sp}\sim (g_s M^2 \, y_{\rm UV})^{2/3}$.  The mass
of the conifold modulus $Z$ scales as $m_Z\sim \Phi/(g_s M^2)$ and
the coefficient in the three-point vertex $ \gamma \phi h_n^2$ reads
\eq{
                  \gamma\sim m(\Phi) \partial_\Phi m(\Phi) \sim
                  {\Phi\over (g_s M^2\, y_{\rm UV})^2} \ll 1\,
}
so that perturbation theory makes sense. We notice that, in
contrast to the SDC for infinite field
distances, at the conifold point $\Delta m$ does not scale
exponentially with the proper field distance but only linearly.
In addition, the number of light species does not increase
exponentially but stays constant. The differences between the two cases are summarized in Figure \ref{fig:ModuliSpace}.
\noindent
As indicated the warped KKLT scenario lies in the blue region
  where KK modes are lighter than the cut-off scale. In this respect,
  KKLT is analogous to e.g. large field inflationary models, that
  require trans-Planckian field distances.

\subsubsection*{Remarks}

Let us close this section with two remarks.
First, we compare the energy density of the AdS minimum \eqref{AdScosmo}
with the cut-off scale
\eq{
       {|V_0|\over \Lambda^4}\sim {1\over g_s M^2}
       \left({|Z|\over {\cal V}_w}\right)^{2\over 3}\sim {m_Z^2\over
         M_{\rm pl}^2} \ll 1\,.
}
Therefore  self-consistently the vacuum energy in the AdS minimum is below the cut-off.

Second, we have seen at the end of section \ref{sec_KK} that massive string
excitations  also give rise to  KK modes localized in the warped
throat with red-shifted masses. Clearly, we have ignored these
potentially ultra-light states in our former analysis. Requiring that
their
off-set in mass \eqref{redshiftstring} is larger than the species
scale leads to $y_{\rm UV}\gtrsim { M^{5/2}/g_s^{1/4}}$.
In case that  the length $y_{\rm UV}$ of the throat is smaller, 
it would be interesting to study the effect of integrating out also
these extra light modes. This more involved analysis is beyond the
scope of this paper.

\subsection{The meaning of the cut-off}

The question now arises whether this so determined cut-off $\Lambda$ 
has any intrinsic meaning. We will show next that it corresponds to the mass of the non-perturbative state given by the wrapped D3-brane on the $S^3$ of the warped deformed conifold. The unwarped case was discussed in section \ref{sec_two-one}.

\subsubsection*{Light wrapped D3-brane}

Prior to the orientifold projection, the complex structure modulus $Z$
was part of a whole ${\cal N}=2$ vectormultiplet, where the vector
field arises from the dimensional reduction of the R-R four-form along
the conifold $A$ cycle with the topology of a three-sphere.  The
corresponding electrically charged object is given by a D3-brane wrapping this $S^3$.
Whether this vector field is projected out or not depends on the
orientifold projection. In the following we assume that both the
complex field $Z$ and the vector field survive the projection\footnote{This
  can e.g. be achieved by letting the holomorphic involution $\sigma$ in $\Omega\sigma (-1)^{F_L}$
act such that it exchanges two such conifold $A\leftrightarrow A'$ cycles. In this
way $A-A'$ will support the complex modulus $Z$ and $A+A'$ the vector field.
In spite of  the non-vanishing flux $\int_{A} F_3=-\int_{A'} F_3=K$
there will be no (S-dual) Freed-Witten anomaly along the 3-cycle $A+A'$
wrapped by the D3-brane.}.
The lightest  D3-brane will be the one wrapping  the three-cycle $S^3$
at the bottom of the throat of the deformed conifold.
In order to  estimate  its mass,
we start with the D3-brane action and dimensionally
reduce it as
\eq{
         S_{\rm D3}\sim {M_s^4\over g_s} \int dt \int_{S^3} d^3 y \sqrt{-G}\sim  
                 {M_s^4\over g_s} \int dt \int_{S^3} d^3 y \, e^{-2A} \sqrt{ \tilde
                   g_{\rm CY} }\,.
}
Using the scaling of the metric \eqref{KSmetric} and the warp factor \eqref{warpfactor2},
we obtain for  the mass
\eq{
\label{masswrappedd3}
                      m^2_{\rm D3}\sim g_s^{1\over 2} M^2  ({\cal V}_w
                      |Z|^2)^{1\over  3} M_s^2\sim
              g_s M^2  \left( {|Z|\over {\cal V}_w}\right)^{\!{2\over  3}} M_{\rm pl}^2\,,
}
which scales precisely as the cut-off $\Lambda$ determined in the
previous section.

Moreover, the mass of this non-perturbative state, with respect to
$({\cal V}_w  |Z|^2)$,  scales in the same way as for the conifold modulus
$Z$ and the lightest KK-modes. However, in the regime $g_s M^2\gg 1$ it
is still heavier than the latter. 
Recall that this was also the regime where an uplift via ${\ov{D3}}$-branes could work.

Recall that in the unwarped case the singularity in the field space
metric
arose from integrating out the corresponding wrapped D3-brane.
Let us discuss whether this also happens in the warped case.
Integrating out the chiral supermultiplet corresponding to the wrapped D3-brane of mass \eqref{masswrappedd3}
leads to the one-loop correction 
\eq{
           g^{\rm 1-loop, D3}_{Z\ovmod Z}&\sim \Big(\partial_{|Z|} m_{\rm
             D3} \Big)^2 \left(1+\alpha \log\bigg({\Lambda^2\over
               m^2_{\rm D3}}\bigg) \right)\\[0.2cm]
&\sim   {g_s M^2\over  ({\cal V}_w
                      |Z|^2)^{2\over  3} } \left(1+\alpha \log\bigg({\Lambda^2\over
               m^2_{\rm D3}}\bigg) \right)\,
}
where $\Lambda$ denotes the cut-off of the effective theory that
includes the wrapped D3-brane. This is not known and only if it satisfies
$\Lambda=c\, m_{D3}$ with a numerical factor $c>1$ one really gets a one-loop correction
that is proportional to the tree-level metric \eqref{hannover96}.
If instead $\Lambda=M_{\rm pl}$ (as in the unwarped case) then the
functional form of the one-loop correction does not match the
tree-level metric. In this case, the singularity in the field space metric
would not emerge from integrating out the non-perturbative D3-brane,
but rather could be interpreted as arising from integrating out the tower of light KK-modes. The
wrapped D3-brane would instead fix the cut-off of the effective theory for
the ultra-light red-shifted modes.

\subsubsection*{Connection to the length of the throat}

As we have seen, for the Wilsonian effective theory of the K\"ahler and the
Z-modulus  one expects the ultimate cut-off to be  
\eq{
    \Lambda= m_{\rm D3}\sim \sqrt{g_s M^2} \left({|Z|\over {\cal
          V}_w}\right)^{\!{1\over 3}} M_{\rm pl}\,
}
and not $M_{\rm pl}$. In section \ref{sec_22} we introduced the length cut-off  $y_{\rm UV}$ of
  the warped throat. This is the location where the KS throat ends
  into a bulk Calabi-Yau threefold, where warping becomes small.
The question is whether this cut-off has anything to do with the
energy cut-off $\Lambda$ that we found for the validity of the
effective action.  

To derive such a relation, let us compute the contribution of the
warped region  to the warped volume ${\cal V}_w$. 
\eq{
\label{warpvola}
     {\cal V}^{\rm throat}_w&={1\over (\alpha')^3 g_s^{3\over 2}} \int d^6 y \sqrt{\tilde
       g_{\rm CY}}\, e^{-4A} \\
    &\sim {1\over (\alpha')^3 g_s^{3\over 2}} \bigg( (\alpha')^3
      g_s^{3\over 2} {\cal V}_w |Z|^2\bigg) \bigg( {g_s M^2\over ({\cal
          V}_w |Z|^2)^{2\over 3}}\bigg) \int_0^{y_{\rm UV}}\! dy\,
      \sinh^2(y) I(y) \\
   &\sim {\cal V}_w \bigg( g_s M^2 \left({|Z|\over {\cal
         V}_w}\right)^{2\over 3} \bigg) \int_0^{y_{\rm UV}}\! dy\,
      \sinh^2(y) I(y) 
}
We note that the combination in the  bracket of the right hand
side in \eqref{warpvola} is precisely $(\Lambda/M_{\rm pl})^2$.
For self-consistency ${\cal V}^{\rm throat}_w$  must be smaller than the total warped volume
${\cal V}_w={\cal V}^{\rm bulk}_w+{\cal V}^{\rm throat}_w$. This
yields a contraint on the length of the throat $y_{\rm UV}$
\eq{
        \int_0^{y_{\rm UV}}\! dy\,
      \sinh^2(y) I(y) \lesssim \left( {M_{\rm pl}\over \Lambda}\right)^2\,.
}
For large values $y_{\rm UV}\gg 1$ we can approximate
\eq{
        \int_0^{y_{\rm UV}}\! dy\,
      \sinh^2(y) I(y) 
\gtrsim     \exp\left({\textstyle {2\over 3}}y_{\rm UV}\right)\,
}
so that we finally obtain the upper bound 
\eq{
                      y_{\rm UV} \lesssim 3\log\left( {M_{\rm pl}\over \Lambda}\right)\,.
}
This provides  an intriguing relation between the cut-off of the effective
theory and the cut-off length scale of the warped throat.

\subsubsection*{Remark on WGC}

Recall that the magnetic version of the WGC provides information on the cut-off of
a $p$-form gauge theory. For instance for a usual one-form gauge field
with gauge coupling $g$ in four-dimensions one has $\Lambda\sim g\, M_{\rm pl}$.
It would be a nice check if also in our case such a gauge field could
be identified that leads via the magnetic WGC to the cut-off $\Lambda=m_{\rm
  D3}$.

We did not manage to find such a gauge field which we suspect is due
to the following. For both toroidal compactifications and the unwarped
conifold, the gauge fields in question are the ones under which the light
KK/D3-branes are electrically charged. In the first case, this is the
$U(1)$ gauge field that arises from the dimensional reduction of the 
off-diagonal components  of the 10D metric. In the second case, this is the 
$U(1)$ gauge field arising from the dimensional reduction of the R-R
four-form along the conifold $A$-cycle. 

In our case, the light modes
are  KK-modes in the throat $y$-direction and therefore we are looking for a gauge field
arising from the reduction of the metric along this
direction. However,  a CY does not contain
any non-trivial one-cycles so that  such a gauge field does not
exist.

\subsection{Extension of the SDC and emergence}

It is tempting to combine the observations at large distances and at
the conifold locus captured in figure \ref{fig:ModuliSpace}, into an extension of the swampland distance
conjecture. The starting point is an effective action $S$ governing the
dynamics  of some light scalar fields $\phi$ with tree-level metric
$g^{(0)}_{\phi\phi}$. In this moduli space there exist points where a
tower of (KK) modes turn out to become lighter than the cut-off scale.
Extending the effective theory to also include this tower of states
will define a new action $\hat S$, which features   a new tree-level metric on
moduli space $\hat g^{(0)}_{\phi\phi}$, about which not much is known.
The essential observation is about the one-loop correction induced by
integrating out the tower of light states leading us to propose:

\begin{quotation}
\noindent
{\bf Extension to SDC:}
{\it There exist points at finite or infinite distance  in moduli
  space with singular tree-level metric $g^{(0)}_{\phi\phi}$ at which towers of modes become lighter than 
the species scale $\tilde\Lambda_{\rm
    sp}=\Lambda/\sqrt{N}$. Adding these states to the action, they 
  induce a one-loop correction $\hat g^{(1)}_{\phi\phi}$ to the field
  space metric whose functional form  is always proportional to the
  former tree-level metric $g^{(0)}_{\phi\phi}$.}
\end{quotation}

\noindent  The essential question is how this behavior should be
interpreted. As in \cite{Heidenreich:2017sim,Grimm:2018ohb,Heidenreich:2018kpg,Corvilain:2018lgw}, 
one could say that the conifold and infinite distance regime in field space are
emerging from  integrating out the tower of
light states that appear in these regions.
As also discussed in \cite{Corvilain:2018lgw}, one could distinguish  two different ways of emergence.

\vspace{0.2cm}
\noindent
{\bf E1:} The singularity in the former tree-level metric
$g^{(0)}_{\phi\phi}$  is entirely emerging from the one-loop
correction $\hat g^{(1)}_{\phi\phi}$  in the extended effective
theory $\hat S$, i.e. $\hat g^{(0)}_{\phi\phi}=0$.

\vspace{0.2cm}
\noindent
{\bf E2:} 
 In $\hat S$ there also exists a non-vanishing
tree-level metric $\hat g^{(0)}_{\phi\phi}\simeq g^{(0)}_{\phi\phi}$
and it is a peculiar property of effective actions of quantum gravity
that the one-loop correction is also proportional to $g^{(0)}_{\phi\phi}$.

\vspace{0.4cm}
\noindent
The emergence of type {\bf E1} was claimed to underly the singularity 
appearing in the case of the unwarped conifold discussed in section 2.
The $\log$-term in the field space metric of the perturbative string 
was considered to be induced by having integrated out a
non-perturbative state, namely the wrapped D3-brane.
 
The second question is how far the original effective action $S$ with
field space metric $g_{\phi\phi}$ and effective potential $V(\phi)$ can be trusted when one is working
in the regions of the moduli where those towers of states become light. These
were the colored regions around the singular points in figure  \ref{fig:ModuliSpace}.
In this respect we can also imagine two different possibilities that
lead to completely opposite conclusions:

\vspace{0.2cm}
\noindent
{\bf R1:} The  initial Wilsonian effective action $S$ and the minima
of the potential are  not reliable because of the tower of extra modes
that are not included in $S$.

\vspace{0.2cm}
\noindent
{\bf R2:} The peculiar property  $\hat g^{(1)}_{\phi\phi}\simeq
g^{(0)}_{\phi\phi}$ signals that the effective action $\hat S$ is {\it not} completely
out of control. Since the superpotential is not expected to
    be perturbatively corrected, also the effective
    potential in $\hat S$ is only slightly changed from the tree-level form $V(\phi)$ in
    $S$.

\vspace{0.4cm}
\noindent
The interpretation {\bf R1}  is the one  followed in the recent discussions of the
swampland distance conjecture and its application to large field
inflation. The second possibility {\bf R2} says that
the certainly present corrections due to the light (KK) modes 
are essentially harmless, as they just change numerical factors
and the naive effective potential $V(\phi)$ is also valid in the regions
close to the singularities in moduli space.
Thus, the extension of the SDC formulates a surprising, 
sort of self-repairing property of any effective theory of quantum
gravity.

Clearly, the implications for the AdS/dS minimum of the warped KKLT construction will
depend on which picture is correct.
In case of {\bf R1}, the warped KKLT construction is
based on a Wilsonian effective action that is not under control
and  the AdS/dS minimum is fake. 
In this respect it is as trustworthy as,  for
instance, the effective action including  non-geometric 
fluxes \cite{Blumenhagen:2015kja} where also KK modes spoiled
the validity of the employed effective action.
Recall that the latter also led to de 
Sitter vacua \cite{Danielsson:2012by,Blaback:2013ht}.

In case of {\bf R2}, though we are working with the naive action
$S$, the scalar potential also gives reliable results in the region
close to the conifold singularity and  the AdS/dS minima
have a good chance to survive in the action $\hat S$. Whether this is
indeed the case requires further studies.

\subsubsection*{Remarks}

Let us close this section with two remarks.
For the original KKLT scenario \cite{Kachru:2003aw} with an assumed tuning of $0<|W_0|\ll 1$ in
the string flux landscape, analogously  the dS uplift does not work in a
controlled way. However, the scale separated AdS minimum could also be found
by using the reliable non-warped effective action in the dilute flux
limit, as reviewed  in section  \ref{sec_two-one}.
Therefore, either this  provides   indeed a counter example to the 
AdS scale separation
conjecture or there is something wrong with this setup. A  reasonable guess is  
 the assumption of  a possible landscape tuning 
$0<|W_0|\ll 1$ in  controlled flux compacifications.
Maybe, all controlled effective values of $W_0$ are larger than an order one number. 
Such  an observation for the distribution of flux vacua for a
concrete Calabi-Yau threefold
was reported  in \cite{MartinezPedrera:2012rs}, but this issue
deserves further studies. Moreover, our findings will also affect all other
global string theory constructions where warped throats are employed,
as for example the dS uplift in the Large Volume Scenario
\cite{Balasubramanian:2005zx}.

Secondly, we remark that in \cite{Ooguri:2018wrx} a derivation of a refined dS swampland
conjecture from the swampland distance conjecture was presented.
Going through the steps of this derivation, as the authors claim
themselves, it is assumed that one is  working in a large field regime
where the spacing of the tower of light  modes and their number 
scales exponentially. Therefore, their derivation does not directly
apply to the regime close the conifold point and dS minima in its vicinity are not
immediately excluded by their arguments.

\section{Conclusions}
\label{sec_concl}

In this paper we have continued the analysis of an effective action that is supposed to be
valid in the strongly warped regime close to a conifold singularity 
of  a type IIB compactification on a CY threefold. 
This is the regime which is relevant for describing the uplift in the
KKLT scenario.
The difference to former
studies is that the dynamics of the conifold
modulus was seriously taken into account, motivating us to propose a
modified
version of the KKLT scenario where initially we assume that all
remaining
complex structure moduli are stabilized supersymmetrically with
$W_{\rm cs}=0$. The effective action for the two moduli system of the
conifold modulus and the overall K\"ahler moduli just by itself
realize the KKLT scenario with an effective exponentially small $W_0$
being generated dynamically.  In addition,  an uplift to de Sitter is
achievable by the introduction of anti D3-branes. 

\begin{figure}[ht]
  \centering
  \includegraphics[width=0.45\textwidth]{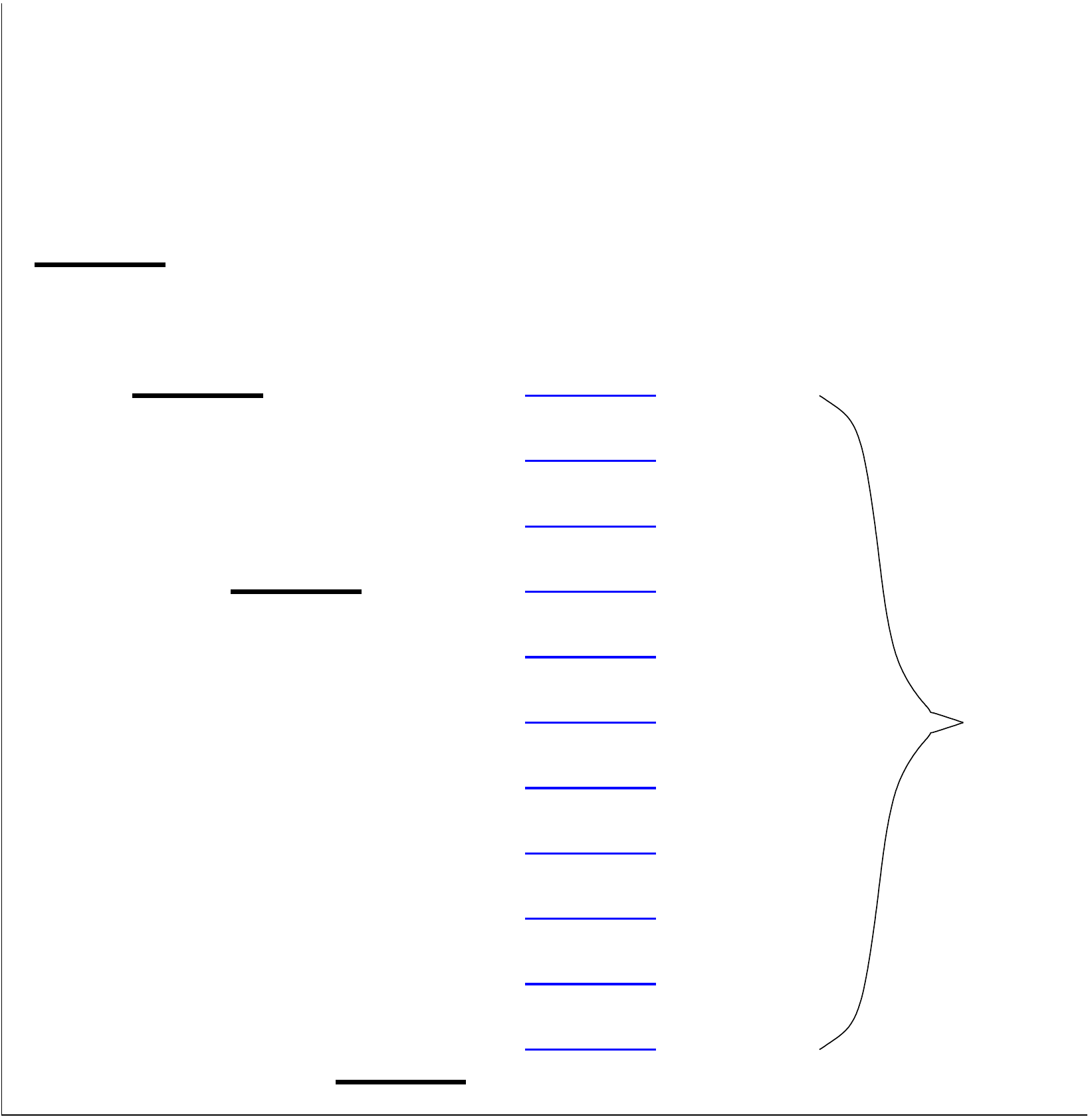}
\begin{picture}(0,0)
    \put(-157,141){$\Lambda=m_{D3}$}
        \put(-140,119){$\tilde\Lambda$}
        \put(-171,87){$m_{Z}$}
        \put(-153,5){$m_{T}$}
        \put(-20,64){KK-modes}
        \put(-235,180){Energy}
        \put(-189,192){\vector(0,1){0}}
    \end{picture}
  \caption{Mass scales in the warped throat in the regime $y_{\rm UV}>1$.}
  \label{fig:spec}
\end{figure}
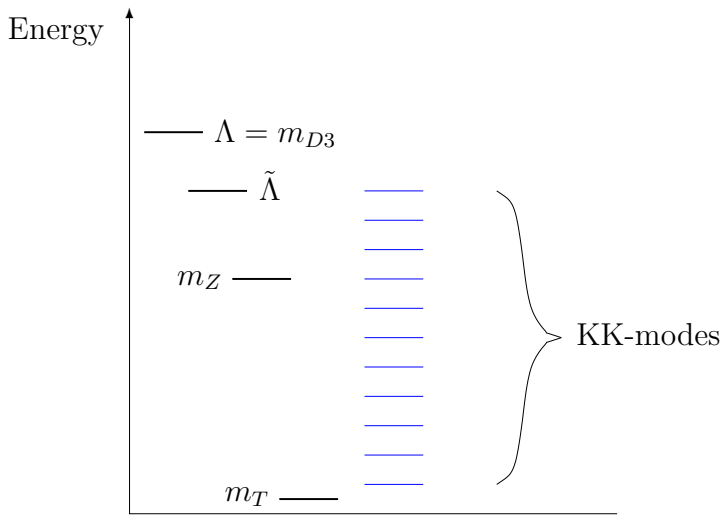

As opposed to the initially fixed 
complex structure moduli, the mass of the conifold modulus comes
out exponentially small, though still larger than the mass of the
K\"ahler modulus. Having available an explicit expression for  the
mass of the conifold modulus, we were comparing it  to masses of
Kaluza-Klein modes localized in the strongly warped throat.
In the supergravity regime
$g_s |M|\gg 1$ and  $g_s |M| y^2_{\rm UV} \gg 1$  there exists a finite number of KK modes that are lighter
than the natural UV cut-off of the effective theory. The latter turned
out not to be the Planck-scale but the species scale related to the mass scale of the non-perturbative state of a
D3-brane wrapped around the three-cycle that shrinks to a point at the conifold
locus. This mass spectrum is shown in figure \ref{fig:spec}.

%%%%%%%%%%%%
%%%%%%%%%%%%
%%%%%%%%%%%%
%%%%%%%%%%%%

The behavior in this regime close to the conifold locus is consistent with the logic
recently employed for the emergence of the swampland distance conjecture in the large
distance regimes. However, some of the details  are  different.
For the conifold, the cut-off was also field dependent leading to two
effects
\begin{itemize}
\item{the spacing of the KK modes scale linearly (and not
    exponentially) with the proper field distance}
\item{the number of light species with a mass below the species scale 
        is constant and not exponentially large\,.}
\end{itemize}

\noindent
Integrating out the finite number of light KK modes leads to a one-loop
correction to the K\"ahler potential that is proportional to the
tree-level one leading us to formulate an extension of the RSDC.
Finally, we were discussing the implications for the 
KKLT AdS/dS minima. Either, the KKLT construction is
based on a Wilsonian effective action that is not under control or
the certainly present corrections due to the light KK modes 
are essentially harmless, as they just change numerical factors.

It is beyond the scope of this paper to fully clarify this issue, 
as this would require a much more thorough analysis of the 
effective action combining the moduli $Z,T$ with the finite tower
of KK modes. Here we just made an attempt to sharpen some relevant,
not yet emphasized issues that appear in the KKLT construction and to connect them
to recent discussions of swampland conjectures.

\vspace{0.5cm}

\noindent
\subsubsection*{Acknowledgments}
We would like to thank Max Brinkmann, Severin L\"ust, Andriana Makridou and Eran Palti for discussions.

\clearpage
%\appendix

%%%%%%%%%%%%%%%%%%%%%%%%%%%%%%%%%%%%%%%%%%%%%%%
%%%%%%%%%%%%%%%%%%%%%%%%%%%%%%%%%%%%%%%%%%%%%%%
%%%%%%%%%%%%%%%%%%%%%%%%%%%%%%%%%%%%%%%%%%%%%%%
%%%%%%%%%%%%%%%%%%%%%%%%%%%%%%%%%%%%%%%%%%%%%%%
%%%%%%%%%%%%%%%%%%%%%%%%%%%%%%%%%%%%%%%%%%%%%%%
%%%%%%%%%%%%%%%%%%%%%%%%%%%%%%%%%%%%%%%%%%%%%%%  
%%%%%%%%%%%%%%%%%%%%%%%%%%%%%%%%%%%%%%%%%%%%%%%
%%%%%%%%%%%%%%%%%%%%%%%%%%%%%%%%%%%%%%%%%%%%%%%

%%%%%%%%%%%%%%%%%%%%%%%%%%%%%%%%%%%%%%%%%%%%%%%
%%%%%%%%%%%%%%%%%%%%%%%%%%%%%%%%%%%%%%%%%%%%%%%
%%%%%%%%%%%%%%%%%%%%%%%%%%%%%%%%%%%%%%%%%%%%%%%
%%%%%%%%%%%%%%%%%%%%%%%%%%%%%%%%%%%%%%%%%%%%%%%

\clearpage
\bibliography{references}  
\bibliographystyle{utphys}

%%%%%%%%%%%%%%%%%%%%%%%%%%%%%%%%%%%%%%%%%%%%%%%
%%%%%%%%%%%%%%%%%%%%%%%%%%%%%%%%%%%%%%%%%%%%%%%
%%%%%%%%%%%%%%%%%%%%%%%%%%%%%%%%%%%%%%%%%%%%%%%
%%%%%%%%%%%%%%%%%%%%%%%%%%%%%%%%%%%%%%%%%%%%%%%

\end{document}